\documentclass[twocolumn]{aastex63}
\newcommand{\kms}{km s$^{-1}$}

\newcommand{\absmue}{$\langle\mu\rangle_{\rm e, abs} (r')$}

\newcommand{\Reffc}{$R_{\rm eff,c}$}
\newcommand{\SBunit}{mag arcsec$^{-2}$}
\newcommand{\kpc}{$\rm kpc$}
\newcommand{\snu}{\affil{Astronomy Program, Department of Physics and Astronomy, Seoul National University, Gwanak-gu, Seoul 08826, Republic of Korea}}

\shorttitle{The Nature of Ultra-diffuse Galaxies in Abell 370}
\shortauthors{Lee et al.}

\begin{document}
	
\title{\textbf{\large{The Nature of Ultra-diffuse Galaxies in Distant Massive Galaxy Clusters: Abell 370 in the Hubble Frontier Fields}}}
\author{Jeong Hwan Lee}\snu
\author{Jisu Kang}\snu
\author{Myung Gyoon Lee}\snu\email{mglee@astro.snu.ac.kr}
\author{and In Sung Jang}\affil{Leibniz-Institut f\"{u}r Astrophysik Potsdam (AIP), An der Sternwarte 16, D-14482, Potsdam, Germany}

\begin{abstract}
We report the discovery of ultra-diffuse galaxies (UDGs) in Abell 370 ($z=0.375$).
We find 46 UDGs in Abell 370 from the images of the Hubble Frontier Fields (HFF).
Most UDGs are low-luminosity red sequence galaxies, while a few of them are blue UDGs.
We estimate the abundance of UDGs in Abell 370, $N(\rm UDG)=644\pm104$.
Combining these results with those of Abell S1063 ($z=0.348$) and Abell 2744 ($z=0.308$) \citep{Lee17}, we derive a mean radial number density profile of UDGs in the three clusters.
The number density profiles of UDGs and bright galaxies show a discrepancy in the central region of the clusters: the profile of UDGs shows a flattening as clustercentric distance decreases, while that of bright galaxies shows a continuous increase.
This implies that UDGs are prone to disruption in the central region of the clusters.
The relation between the abundance of UDGs and virial masses of their host systems is described by a power-law with an index nearly one: $N({\rm UDG})\propto M_{200}^{0.99\pm0.05}$ for $M_{200}>10^{13}~M_{\odot}$.
We estimate approximately dynamical masses of UDGs using the fundamental manifold method, and find that most UDGs have dwarf-like masses $(M_{200}<10^{11}$ $M_{\odot})$.
This implies that most UDGs have a dwarf-like origin and a small number of them could be failed $L^{*}$ galaxies.
These results suggest that multiple origins may contribute to the formation and evolution of UDGs in massive galaxy clusters.

\end{abstract}

\keywords{galaxies: clusters: individual 
	(Abell 370)
	--- galaxies: dwarf
	--- galaxies: formation
	--- galaxies: evolution} 

\section{Introduction}

In the extensive photographic studies of the Virgo cluster, \citet{san84} discovered a new class of dwarf galaxies which have very large diameter ($\sim$10 kpc) and low central surface brightness ($\mu_{0}(B)>25$ \SBunit).
They found about 20 such low surface brightness (LSB) galaxies, which are mainly located in the central region of the Virgo cluster.
Similar LSB galaxies were found in other galaxy clusters later and they were called low-mass cluster galaxies (\citet{con03}, \citet{con18}, and references therein).
\citet{van15} found such large LSB galaxies in the Coma cluster and renamed them ``ultra-diffuse galaxies (UDGs)''.
UDGs have exceptionally large sizes ($R_{\rm eff}>1.5~{\rm kpc}$) but low surface brightness ($\mu_{0}(g)>24.0$ \SBunit).
Thus, UDGs seem to be the extreme case of LSB dwarf galaxies.

UDGs have been found in various environments.
In cluster environments, there are hundreds of UDGs detected in the Coma cluster \citep{kod15, yag16}, the Fornax cluster \citep{cal87, mun15, ven17}, the Virgo cluster \citep{san84, imp88, mih15, mih17}, eight clusters at $z=0.044-0.063$ from Multi-Epoch Nearby Cluster Survey (MENeaCS) \citep{vdB16}, Abell 168 \citep{rom17a}, the Perseus cluster \citep{con03, wit17}, 18 clusters from MENeaCS \citep{sif18}, 8 clusters from the Kapteyn IAC WEAVE INT Clusters Survey (KIWICS) \citep{man18, man19}, Abell S1063 \citep{Lee17}, and Abell 2744 \citep{jan17, Lee17}.
In group environments, UDGs were found in the NGC 5485 group \citep{mer16}, Hickson Compact Groups (HCGs) \citep{rom17b, shi17}, galaxy groups from the KiDS and GAMA fields \citep{vdB17}, the Leo-I group \citep{mul18}, and galaxy groups from the Dragonfly Nearby Galaxies Survey \citep{coh18}.
Isolated UDGs are rare, but some have been found in wide field surveys: DGSAT-I \citep{mar16}, UGC 2162 \citep{tru17}, SdI-1 and SdI-2 \citep{bel17}, HI-bearing ultra-diffuse sources (HUDS) \citep{lei17}, low surface brightness galaxies (LSBGs) from Hyper Suprime-Cam Subaru Strategic Program (HSC-SSP) \citep{gre18a}, and S82-DG-1 \citep{rom19}.

The properties of UDGs vary with their environments.
Cluster UDGs generally have red colors ($g-i\sim0.8$), round shapes (axis ratios $b/a\sim0.8$), smooth exponential light profiles (S\`ersic indices $n\sim1.0$), and gas-deficient properties \citep{kod15, van15}.
Thus, UDGs in high density environments are composed of old stellar populations, and their star formation is considered to have been quenched at an early time by gas removal processes.
In contrast, most UDGs in low density environments have bluer colors ($g-i<0.5$), irregular shapes with star-forming knots, and relatively high HI masses ($M_{\rm HI}\sim10^{8}$ $M_{\odot}$) \citep{rom17b, spek18}.
Therefore, these isolated UDGs have young stellar populations and gas-rich properties.
The existence of these various types of UDGs in different environments continues to give us questions about the origin of UDGs.

A key motivation of the studies of UDGs is their formation mechanisms.
There are three main scenarios proposed to explain the formation of UDGs in the literature (\citet{amo16, van16, dic17, ben18} and references therein).
In the first scenario, the so-called ``failed galaxies'' scenario, UDGs failed to generate a typical amount of stars given their halo masses due to environmental effects.
In the second scenario, the ``extended dwarf galaxies'' scenario, UDGs were extended from normal dwarf galaxies due to internal processes.
In the third scenario, UDG progenitors were tidally extended by interactions with neighboring massive galaxies.
Since diverse types of UDGs have been found in observations, recent studies have mostly suggested mixed formation mechanisms of UDGs with a combination of the above scenarios \citep{Lee17, pap17, ala18, FeM18, Lim18, pan18}.

In this work, we search for UDGs in the massive galaxy cluster Abell 370 ($z=0.375$) in the Hubble Frontier Fields (HFF) \citep{Lot17}, following our previous work on UDGs in Abell S1063 and Abell 2744 \citep{Lee17}.
We use very deep HFF archival images to search for UDGs in the clusters.
We combine the results of this work and the previous work of Abell S1063 and Abell 2744 to understand the nature of UDGs in massive galaxy clusters.
After submitting the first draft of our paper, \citet{jan19} presented a study of UDGs and ultra compact galaxies in six HFF clusters including Abell 370.
In this study, we cover not only UDGs but also LSB dwarfs in the clusters.

This paper is organized as follows.
We describe the data from the HFF and our data analysis in \textbf{Section 2}.
We explain how we determined the depth of the HFF data, performed photometry, and selected UDG samples.
In \textbf{\S3}, we show our main results: color-magnitude diagrams (CMDs), color-color diagrams (CCDs), distribution of structural parameters of the UDGs, radial number density profiles (RDPs) of the UDGs, the abundance of UDGs, and estimation of their dynamical masses.
Then, we present implications of the results and discuss the nature of UDGs in \textbf{\S4}.
We present our conclusions and summary in the final section.

We adopt the $\Lambda$CDM cosmological parameters with $H_{0}=73$ $\rm{km}$ $\rm{s^{-1}}$ $\rm{Mpc^{-1}}$, $\Omega_{M}=0.27$, and $\Omega_{\Lambda}=0.73$.
The luminosity distance of Abell 370 for these parameters is $d_{L}=1942$ $\rm Mpc$ and the angular diameter distance is $d_{A}=1028$ $\rm Mpc$.
The virial radius and mass of this cluster are adopted from a lensing analysis using deep and wide-field Suprime-Cam imaging \citep{broad08, Umet11}: $r_{200}=2.55\pm0.11$ $\rm Mpc$ and $M_{200}=(3.03\pm0.37)\times10^{15}$ $M_{\odot}$.
The foreground reddening value toward Abell 370 is $E(B-V)=0.028$ \citep{sch11}.
{\color{blue}\bf {\bf Table 1}} shows the basic physical parameters of the Abell 370 cluster.

\section{Data and Data Analysis}

\subsection{Data}
We obtained deep and high-resolution images of Abell 370 from the HFF archive \citep{Lot17}.
We chose drizzled images of ACS/WFC $F606W(V)$, $F814W(I)$, WFC3/IR $F105W(Y)$, and $F160W(H)$ for our analysis.
The HFF images we used cover two fields with different clustercentric distances ($r/r_{200}$): the central field ($r/r_{200}<0.2$) and the parallel field ($r/r_{200}\sim0.6-0.8$) located at $6'$ southeast of the central field.
We used the areas that both ACS/WFC and WFC3/IR covered: 
$\sim$5.5 $\rm acrmin^{2}$ for the central field, and $\sim$5.0 $\rm arcmin^{2}$ for the parallel field.
The pixel scale of all the drizzled images is $0\farcs03$ per pixel, and the scale for the distance to Abell 370 is $4.984$ $\rm kpc$ $\rm arcsec^{-1}$.

\begin{figure*}
	\centering
	\includegraphics[scale=0.75]{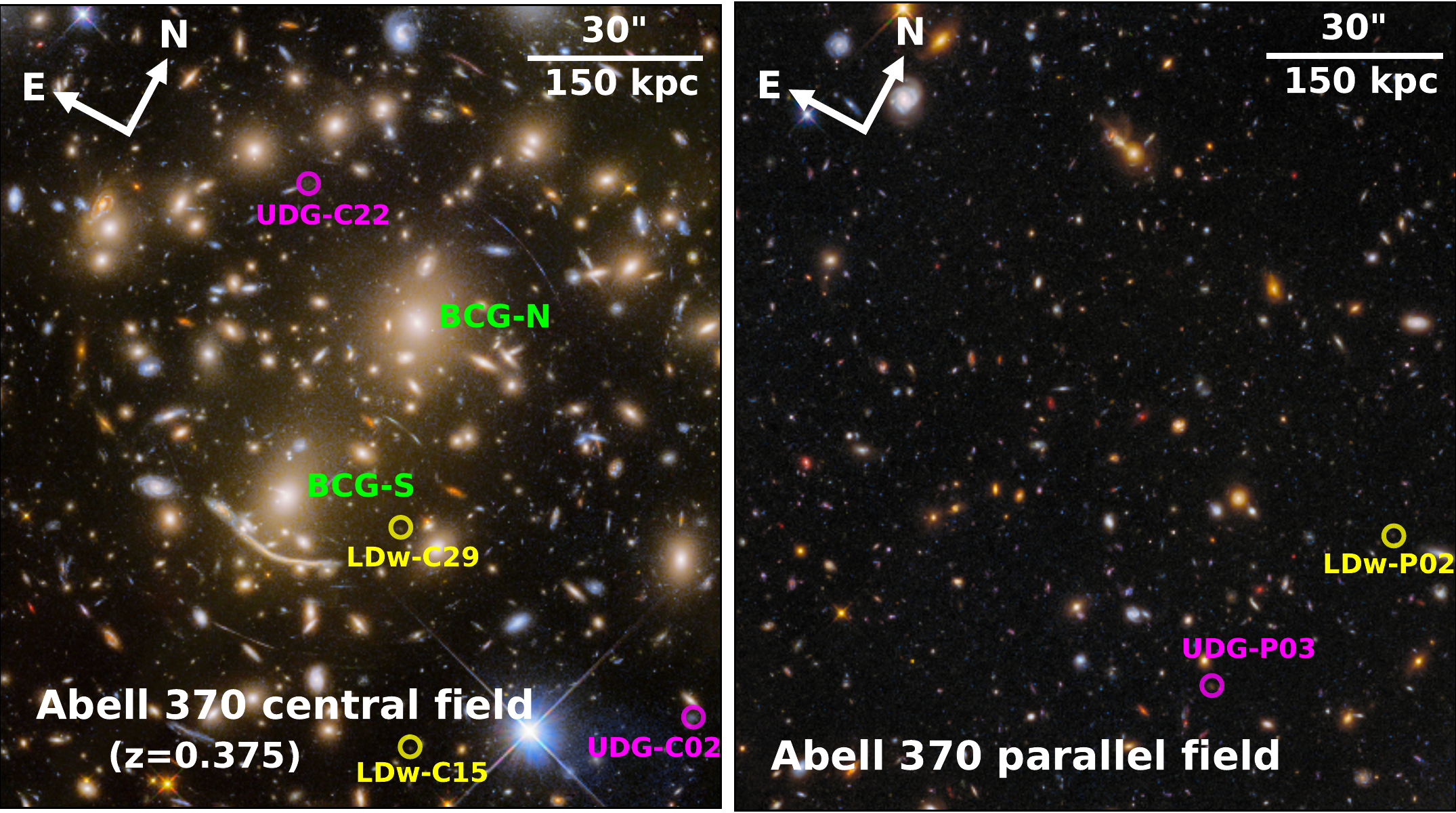}
	\caption{False color images of Abell 370 in the HFF: the central field (left) and the parallel field (right). These images were combined with multi-band images from optical ($F435W$, $F606W$, and $F814W$) to near-infrared ($F105W$, $F125W$, $F140W$, and $F160W$) \citep{Lot17}. The magenta circles are UDG examples and the yellow circles are LSB dwarf examples.
	Two brightest cluster galaxies, BCG-N (north) and BCG-S (south), are labeled in green. The orientation and the image scale  are shown at the top.
	\label{fig_finding}}
\end{figure*}

Among the multi-wavelength bands of the HFF images, $F814W$ and $F105W$ images are the most useful for finding UDGs in Abell 370.
This is because $F814W$ and $F105W$ have the longest exposure times among the available bands, and the spectral energy distributions (SEDs) of cluster galaxies are dominant at these wavelengths considering the cluster redshift ($z=0.375$).
The exposure times for the images of the central field are $36.3$ hours ($F814W$) and $19.3$ hours ($F105W$), and those of the parallel field images are $29.3$ hours ($F814W$) and $20.3$ hours ($F105W$).
The FWHMs of point spread functions (PSFs) are $0\farcs09$ ($\sim$0.45 $\rm kpc$) in $F814W$ and $0\farcs16$ ($\sim$0.80 $\rm kpc$) in $F105W$.
We adopted the foreground extinction values of $A_{F814W} = 0.049$~mag and $A_{F105W}=0.031$~mag given in the HFF-DeepSpace photometric catalogs \citep{shp18} for CMDs and CCDs.

To estimate background contributions, we used ACS/WFC $F814W$ and WFC3/IR $F105W$ images of the Hubble eXtreme Deep Field (XDF) \citep{ill13}.
The total exposure times are $14.1$ hours in $F814W$ and $74.1$ hours in $F105W$.
We chose only the deepest regions of the whole XDF, the HUDF09 and the HUDF12 programs, which cover $\sim5.06$ $\rm arcmin^{2}$.
Thus, the areas for the central field of Abell 370 ($\sim5.50$ $\rm arcmin^{2}$), the parallel field ($\sim5.00$ $\rm arcmin^{2}$), and the XDF ($\sim5.06$ $\rm arcmin^{2}$) are similar.
Galaxies in the XDF have a wide range of redshifts, but we assume all of them have the same redshift as Abell 370 ($z=0.375$) to estimate the background contamination in the sample of Abell 370.

{\color{blue}\bf {\bf Figure \ref{fig_finding}}} shows the color composite images of the central field (left panel) and the parallel field (right panel) of Abell 370.
In the left panel, the central field shows diverse types of galaxies: two brightest cluster galaxies (BCGs) dubbed ``BCG-N'' and ``BCG-S'', a sequence of bright red ellipticals at the north, and a large number of gravitationally lensed galaxies.
Among them, examples of LSB galaxies including UDGs and LSB dwarfs (LDws) (see the text in \textbf{\S2.2}) are marked by magenta and yellow circles.
`C' and `P' in the names of these LSB galaxies mean the central field and the parallel field of Abell 370.
In the right panel, the parallel field seems to have a much lower number density of bright galaxies.
This means that the parallel field has a fewer number of cluster member galaxies than the central field.
We marked some examples of UDGs and LSB dwarfs in the parallel field with circles.

\begin{figure}
	\centering
	\includegraphics[scale=0.75]{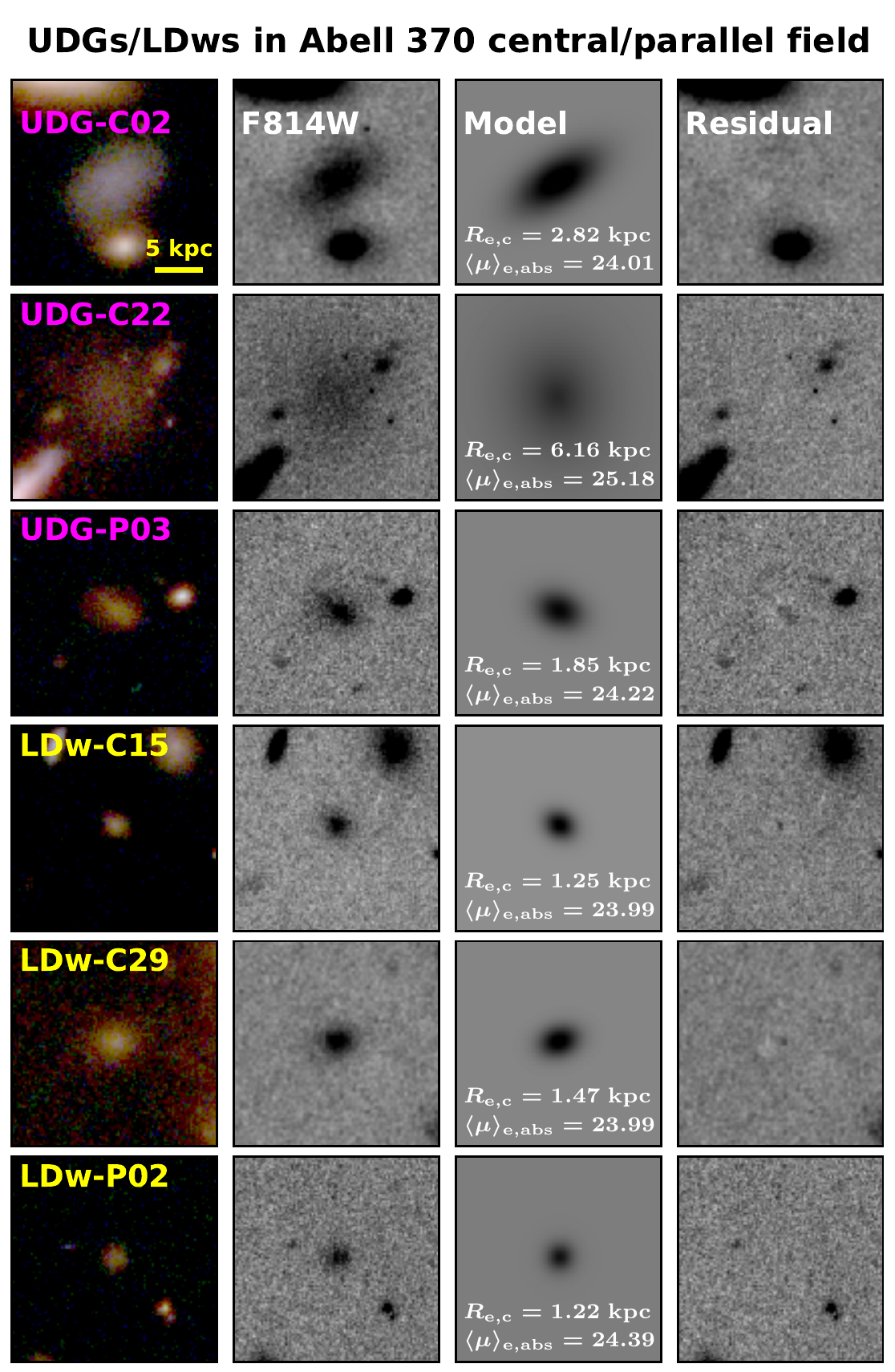}
	\caption{Thumbnail images ($4\farcs5\times4\farcs5$) of UDGs and LSB dwarfs denoted in {\color{blue}\bf {\bf Figure \ref{fig_finding}}}. The first column shows RGB color images of each sample (Blue : $F435W+F606W$ images, Green : $F814W$ images, and Red : $F105W$ images). The second column shows $F814W$ images, which were used as input images for GALFIT. The last two columns show galaxy model images and their residual images from GALFIT measurements. Derived effective radii and surface brightness are marked.
	\label{fig_thumb}}
\end{figure}

{\color{blue}\bf {\bf Figure \ref{fig_thumb}}} shows the zoom-in thumbnail images of the LSB galaxies marked in {\color{blue}\bf {\bf Figure \ref{fig_finding}}}.
The first column is the color images, and the second is $F814W$ band images.
We estimated the sizes and surface brightness of these galaxies using GALFIT version 3.0.5 \citep{pen10} as described in \textbf{\S2.2}.
The third and fourth column show the results from GALFIT, which are galaxy model images and subtracted images, respectively.
The subtracted images show little of original galaxy images, showing that GALFIT did a good job in galaxy modeling.
Most of the LSB galaxies show red colors in their color composite images.
However, one of them (UDG-C02) shows a much bluer color than the others.
We discuss this with our CMDs and CCDs in \textbf{\S 3}.

\begin{figure*}
	\centering
	\includegraphics[scale=0.75]{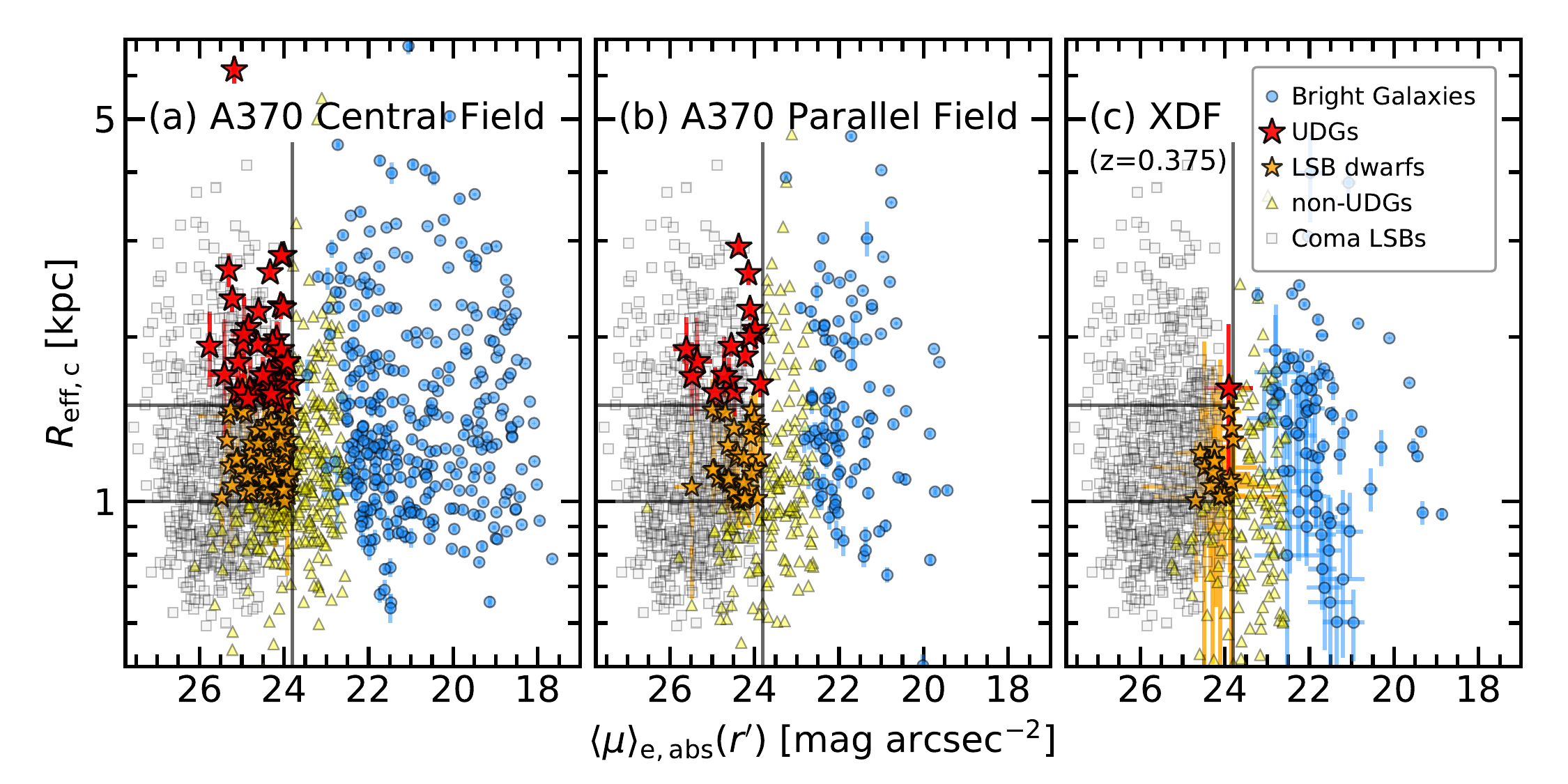}
	\caption{Selection diagrams with absolute mean surface brightness (\absmue) in x-axis and circularized effective radii (\Reffc) in y-axis. Each panel includes all galaxies in the Abell 370 central field (a), the parallel field (b), and the XDF (c). Blue circles show bright galaxies selected with $\mu_{0}(F814W)<22.5$ \SBunit$~$and $R_{\rm eff,c}>1.0$ \kpc. Red and orange stars denote UDG candidates ($\langle\mu\rangle_{\rm e,abs}(r')>23.8$ \SBunit$~$and $R_{\rm eff,c}>1.5$ \kpc) and initial LSB dwarf candidates ($\langle\mu\rangle_{\rm e,abs}(r')>23.8$ \SBunit$~$and $R_{\rm eff,c}=1.0-1.5$ \kpc) selected with visual inspection. Yellow triangles named ``non-UDGs'' do not satisfy any selection criteria of the surface brightness and size. Gray squares are LSB galaxies ($R_{\rm eff} \gtrsim 0.7$ \kpc) discovered in the Coma cluster \citep{yag16}.
	\label{fig_selection}}
\end{figure*}

\subsection{Photometry and Initial Sample Selection}
We performed dual mode photometry using SExtractor \citep{ber96}.
We set $F814W$ images as detection images.
Our configuration parameters for SExtractor photometry are summarized in {\color{blue}\bf {\bf Table \ref{tab_SE}}}.
These parameters are very similar to those used in \citet{Lee17}, which are considered to be effective for  detecting LSB galaxies.
We used the magnitude zeropoints of the AB system for the standard calibration.
In the initial output catalogs of SExtractor, 18,315 sources were extracted in the central field and 15,998 sources in the parallel field.
{\color{blue}\bf {\bf Table 3}} shows the numbers of sources in each step of the sample selection.
Each sample selection step is described below.

First, we selected initial galaxy candidates using the basic parameter criteria: \texttt{MAGERR\_AUTO} $<1.0$, \texttt{FLAGS} $<4$, \texttt{CLASS\_STAR} $<0.4$, and $-0.5<F814W-F105W<1.0$.
This left 3,748 objects in the central field, 2,789 objects in the parallel field, and 2,252 objects in the area of the XDF we used (the HUDF09 and the HUDF12).
Next, we classified these objects into two types of galaxies, bright galaxies and LSB galaxies.
We used the surface brightness and effective radii conditions: bright galaxies with \texttt{MU\_MAX}($F814W$) $<22.5$ \SBunit$~$and \texttt{FLUX\_RADIUS} $>1.0$ \kpc, and LSB galaxies with \texttt{MU\_MAX}($F814W$) $>22.5$ \SBunit$~$, \texttt{FLUX\_RADIUS} $>1.0$ \kpc, and \texttt{B\_IMAGE$/$A\_IMAGE} $>0.3$.
We set the minimum effective radii (\texttt{FLUX\_RADIUS}) to be $1.0$ \kpc, because we intended to select our galaxy samples with effective radii larger than $2\times\rm FWHM$ ($\sim0.90$ $\rm kpc$).
We also set the minimum value of the axis ratio (\texttt{B\_IMAGE$/$A\_IMAGE}) to be 0.3 in order to exclude gravitational lenses and elongated artifacts.
As a result of all these processes, the numbers of selected bright galaxies are 334, 115, and 84 in the central field, the parallel field, and the XDF.
The numbers of selected LSB galaxies are 714, 342, and 339 in each field.

Then, we utilized GALFIT to estimate the surface brightness and  effective radii of the selected galaxies more precisely.
We made input images of each galaxy trimmed with a size of $4\farcs5\times4\farcs5$ in order to estimate background values locally.
This could minimize the effect of diffuse light from bright galaxies in background estimation.
The PSF convolution size is set to be about $5\farcs4$, which is sufficient to give the PSF effect to the entire area of the input images.
Other GALFIT configurations and masking methods are similar to those of \citet{Lee17}.
Most galaxies were fitted by a single S\'ersic law, and nucleated galaxies were fitted with their central nuclei masked.

We divided the sample of LSB galaxies into ``UDGs'' and ``LSB dwarfs''.
We used the selection criteria adopted in \citet{vdB16} and \citet{Lee17}: \absmue$~$$>23.8$$~$\SBunit$~$and \Reffc$~$$>1.5$$~$\kpc$~$for UDGs, and \absmue$~$$>23.8$$~$\SBunit$~$and \Reffc$~$$=1.0-1.5$$~$\kpc$~$ for LSB dwarfs.
In this study, these criteria are used for comparison of the Abell 370 UDGs and the UDGs found in the local universe.
In addition, we selected LSB dwarfs as smaller counterparts of UDGs.
We checked whether UDGs and LSB dwarfs have consistent properties in the result section.
\absmue$~$is the SDSS $r'$-band absolute mean surface brightness at the effective radii, and \Reffc$~$is the circularized effective radii.
We transformed the $F814W$ magnitudes to SDSS $r'$ system using simple stellar population (SSP) models derived from GALAXEV \citep{bc03}.
We adopted an age of 12 Gyr, the Chabrier initial mass function \citep{cha03}, and $[Z/Z_{\odot}]=-0.7$ $(Z=0.004)$ for obtaining SSP models.

Finally, we performed visual inspection of  all selected galaxies to get rid of artifacts, tidal structures, gravitational lenses, and blended sources.
If some galaxies show unreasonable GALFIT results or residuals due to interfering light from neighboring galaxies or saturated stars, we redid manual masking and GALFIT configurations until we obtained reasonable solutions.
The selected galaxies are represented  in the selection diagram of the circularized effective radii vs. the SDSS $r'$-band absolute mean surface brightness at the effective radii, as plotted in {\color{blue}\bf {\bf Figure \ref{fig_selection}}}, and their census is summarized in ``Step 2'' of {\color{blue}\bf {\bf Table 3}}.
Examples of selected UDGs and LSB dwarfs are displayed in {\color{blue}\bf {\bf Figure \ref{fig_thumb}}}.
There are 315, 106, and 80 bright galaxies in the central field, the parallel field, and the XDF.
For UDGs, there are 39, 16, and 1 UDGs in each field.
For LSB dwarfs, there are 87, 35, and 18 sources in each field.
These initial samples are used to select the final samples in \textbf{\S3.1.1}.

\begin{figure}
	\centering
	\includegraphics[scale=0.6]{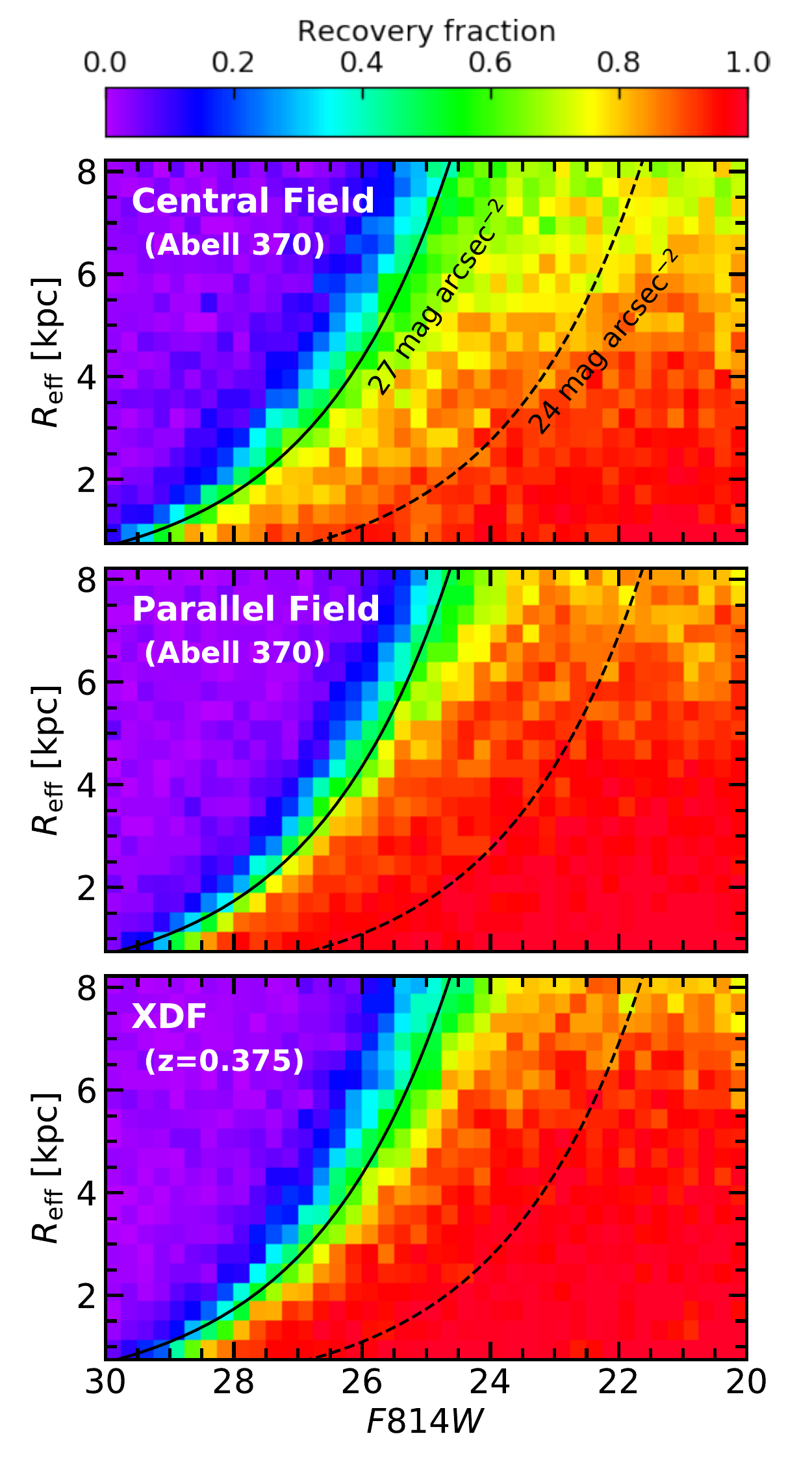}
	\caption{Integrated $F814W$ magnitudes of mock galaxies versus their half-light radii showing the recovery fractions from the artificial galaxy tests of the Abell 370 central field (upper panel), the parallel field (middle panel), and the XDF (the HUDF09 and the HUDF12) (lower panel). Color scale bar at the top denotes the recovery fraction. Black dashed line represents $\langle\mu\rangle_{\rm e,abs}(r')=24~$\SBunit, which is close to the selection criteria for UDGs and LSB dwarfs. Black solid line denotes $\langle\mu\rangle_{\rm e,abs}(r')=27~$\SBunit$~$line, which approximates to the 50\% completeness limits of the two Abell 370 fields and the XDF.
	\label{fig_mocktest}}
\end{figure}

\begin{figure}
	\centering
	\includegraphics[scale=0.425]{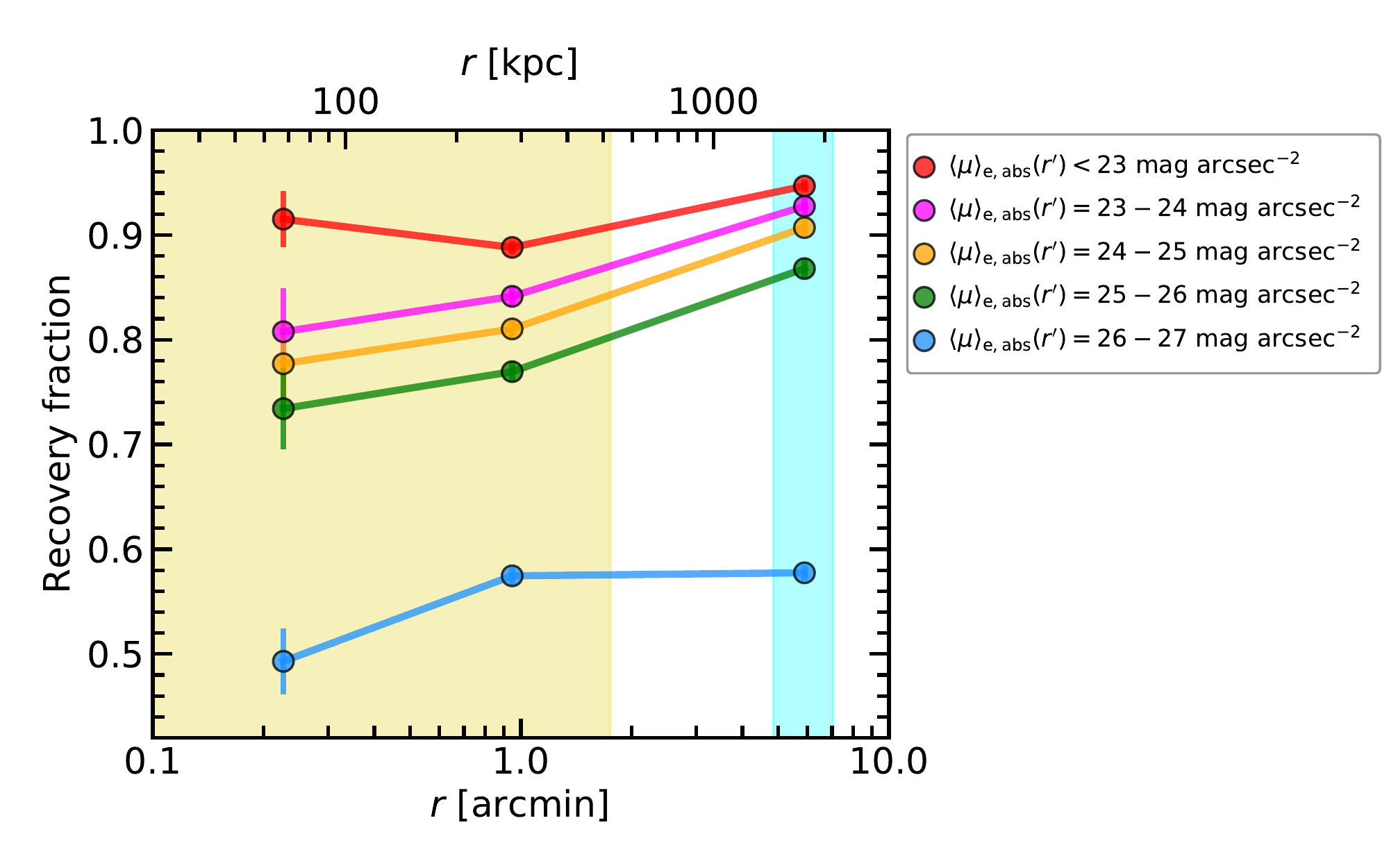}
	\caption{Recovery fractions obtained from artificial galaxy tests as a function of clustercentric distance. Shaded regions in yellow and cyan are the Abell 370 central field and the parallel field. Red, magenta, orange, green, and blue filled circles are for the mock galaxies with $\langle\mu\rangle_{\rm e,abs}(r')<23~$\SBunit, $\langle\mu\rangle_{\rm e,abs}(r')=23-24~$\SBunit, $\langle\mu\rangle_{\rm e,abs}(r')=24-25~$\SBunit, $\langle\mu\rangle_{\rm e,abs}(r')=25-26~$\SBunit, and $\langle\mu\rangle_{\rm e,abs}(r')=26-27~$\SBunit.
	\label{fig_mocktest2}}
\end{figure}

\subsection{Artificial Galaxy Tests}
We carried out artificial galaxy tests for the HFF $F814W$ images of the central field, the parallel field, and the XDF (the HUDF09 and the HUDF12) 
to compute the detection limits and the completeness as a function of clustercentric radius.
Utilizing \textit{artdata/gallist} task in IRAF, we generated a total of 120,000 mock galaxies with uniform distributions of $F814W$ magnitude, size, and spatial number density in each field.
We set 800 bins of $F814W$ magnitude and effective radii: the magnitude bins are from 20 mag to 30 mag with an interval of 0.25 mag, and the effective radii bins are from 0.75 kpc to 8.22 kpc with a linear interval of 0.37 kpc.
There are 150 mock galaxies in each magnitude-size bin.
The axis ratios of mock galaxies are set to follow a Gaussian distribution with a mean value of $\langle b/a\rangle=0.75$, which is the mean value of the axis ratios of Coma LSB galaxies \citep{yag16}.

{\color{blue}\bf {\bf Figure \ref{fig_mocktest}}} shows the results of artificial galaxy tests.
The 50\% surface brightness limit, which is about 27 \SBunit$~$in the absolute SDSS $r'$ magnitude, is similar in all the three fields.
However, it appears that the central field of Abell 370 has slightly lower completeness values for detecting LSB galaxies than the parallel field and the XDF.
{\color{blue}\bf {\bf Figure \ref{fig_mocktest2}}} displays the calculated radial completeness as a function of clustercentric distance.
All the magnitude bins show lower completeness values at the Abell 370 central field than the parallel field.
This is because of diffuse light and a high number density of bright galaxies in the cluster central field.
These completeness values in each radial bin are applied to compute the RDPs of galaxies in \textbf{\S3.4}.

\begin{figure*}
	\centering
	\includegraphics[scale=0.75]{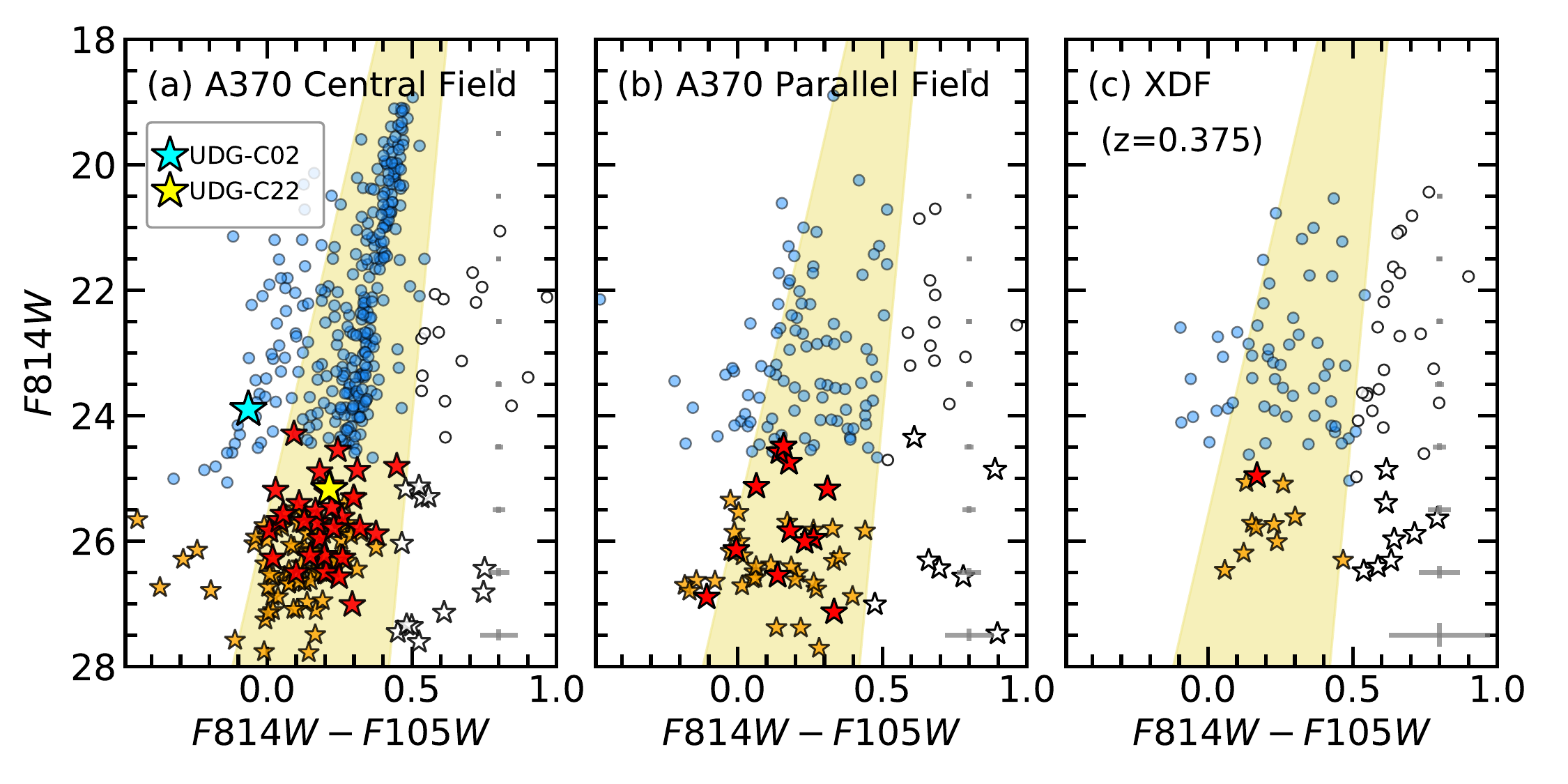}
	\caption{Color-magnitude diagrams (CMDs)
	of galaxies in the Abell 370 central field (a), the parallel field (b), and the XDF (c). Symbols are the same as {\color{blue}\bf {\bf Figure \ref{fig_selection}}}. Gray error bars on the right side of each panel indicate the mean errors of colors and magnitudes for given magnitudes. Yellow shaded regions denote the red sequence of galaxies in the cluster central field. White open symbols mark galaxies excluded from our final samples, because they are redder than the red sequence. 
	Cyan star (UDG-C02) denotes the bluest UDG ($F814W-F105W=-0.07$), and yellow star (UDG-C22) is the largest UDG ($R_{\rm eff,c}=6.16$ \kpc) in our UDG sample.
	\label{fig_cmd}}
\end{figure*}

\section{Results}

\subsection{CMDs of the Galaxies}

\subsubsection{Final Sample Selection}

{\color{blue}\bf {\bf Figure \ref{fig_cmd}}} shows the CMDs of the initially selected galaxies (\textbf{\S2.2}) in the central field of Abell 370 (the left panel), its parallel field (the middle panel), and the XDF assumed to be at the same redshift of Abell 370 (the right panel).
The most remarkable feature in this figure is the red sequence of galaxies in the central field.
We set a boundary of the red sequence as shown by yellow shaded regions in the figure.
There are a small number of galaxies redder than the boundary of the red sequence in each field.
These galaxies are considered to be background galaxies. 
We removed these background galaxies from the initial sample, and produced the final sample of galaxies.

As a result, we identified 298, 93, and 56 bright galaxies in the three fields, 34, 12, and 1 UDGs in each field, and 80, 32, and 10 LSB dwarfs in the final sample.
These numbers are summarized in ``Step 3'' of {\color{blue}\bf {\bf Table 3}}.
In total, we detect 46 UDGs and 112 LSB dwarfs in the central and parallel field of Abell 370.
\citet{jan19} found 65 UDGs in both fields of Abell 370, which is 19 larger than the number in this study.
The reason for this difference is not clear.
It is noted that our study excluded 9 UDG candidates which are much redder than the red sequence, while it is not clear whether \citet{jan19} applied this selection criterion.
For the XDF, the numbers of bright galaxies (56), UDGs (1), and LSB dwarfs (10) are useful to estimate the contribution of background galaxies in the central and parallel fields.
UDGs detected in the central and parallel fields are mostly real members of Abell 370, because the corresponding number of UDGs in the background field is only one.
For the UDGs and LSB dwarfs in the central and parallel fields, we provide the catalogs of their photometric properties ($F814W$ magnitudes, $F814W-F105W$ colors, effective radii \Reffc, effective surface brightness \absmue, S\'ersic indices $n$, and axis ratios $b/a$) in {\color{blue}\bf {\bf Table \ref{tab_UDG}}} and {\color{blue}\bf {\bf Table \ref{tab_LDw}}}.

\begin{figure*}
	\centering
	\includegraphics[scale=0.75]{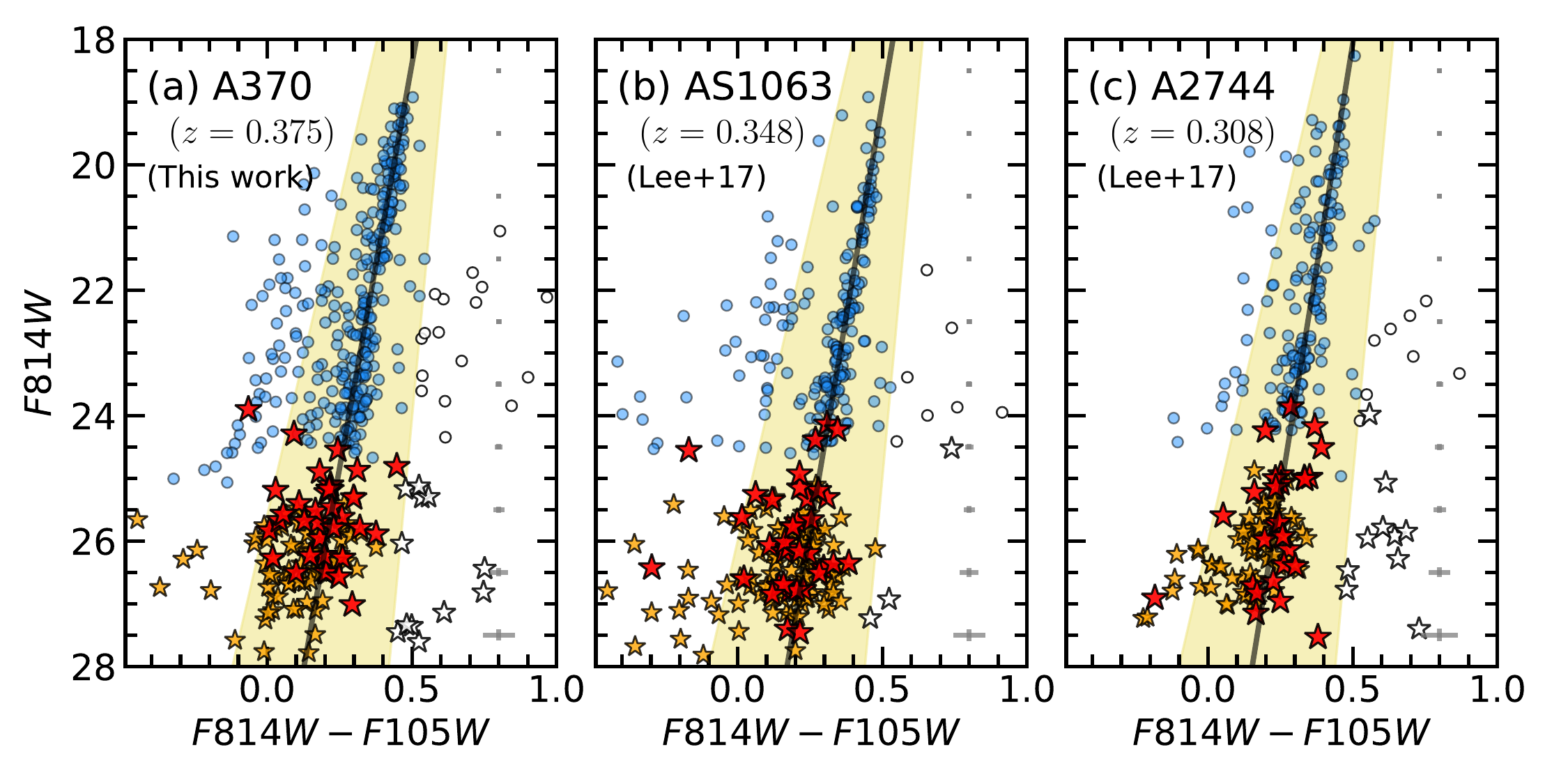}
	\caption{CMDs of galaxies in three HFF clusters: Abell 370 ($z=0.375$), Abell S1063 ($z=0.348$), and Abell 2744 ($z=0.308$). Symbols are the same as {\color{blue}\bf {\bf Figure \ref{fig_selection}}}. Data for Abell S1063 and Abell 2744 are from \citet{Lee17}.
    Black solid lines denote linear fitting lines of the red sequences derived from median colors and magnitudes of galaxies brighter than $F814W<23.5$ $\rm mag$.
    \label{fig_cmd3}}
\end{figure*}

\subsubsection{CMDs of Abell 370}

In {\color{blue}\bf {\bf Figure \ref{fig_cmd}}},  most UDGs in the central and parallel fields of Abell 370 are located at the faint end of the red sequence.
Most LSB dwarfs also have colors similar to those of  UDGs, but a small number of them show bluer colors.
This indicates that most UDGs and LSB dwarfs are made up of an old stellar population, except for a few blue LSB galaxies.

We marked two unusual UDGs in the CMD of the central field.
First, UDG-C02 (cyan star) is the bluest UDG of our sample ($F814W-F105W=-0.07\pm0.01$).
The blue color indicates that this UDG is mainly composed of young stars.
This is in strong contrast to the fact that most of cluster UDGs are located in the red sequence.
Thus, this provides a rare sample of a blue UDG in a massive galaxy cluster.
It is noted that a few blue and irregular UDGs were discovered in various environments \citep{Lee17, rom17b, tru17}, but they are much fainter than UDG-C02 with $M_{r'}=-16.8$.
Second, UDG-C22 (yellow star) is the largest UDG of our UDG sample with \Reffc$~=6.16\pm0.36~$\kpc$~$and \absmue$~=25.18\pm0.08~$\SBunit.
This UDG is significantly larger than the other UDGs in Abell 370 which have \Reffc$~=1.5-3.0$ kpc ($\langle R_{\rm eff,c}\rangle=2.0$ kpc).
This UDG is located right at the red sequence.
UDG-C22 is one of the largest UDGs among the known UDGs including Coma DF17 ($R_{\rm eff}=4.4$ kpc), Coma DF44 ($R_{\rm eff}=4.6$ kpc), and a few Virgo UDGs ($R_{\rm eff}=2.9-9.7$ kpc) \citep{mih15}.

\begin{figure}
	\centering
	\includegraphics[scale=0.475]{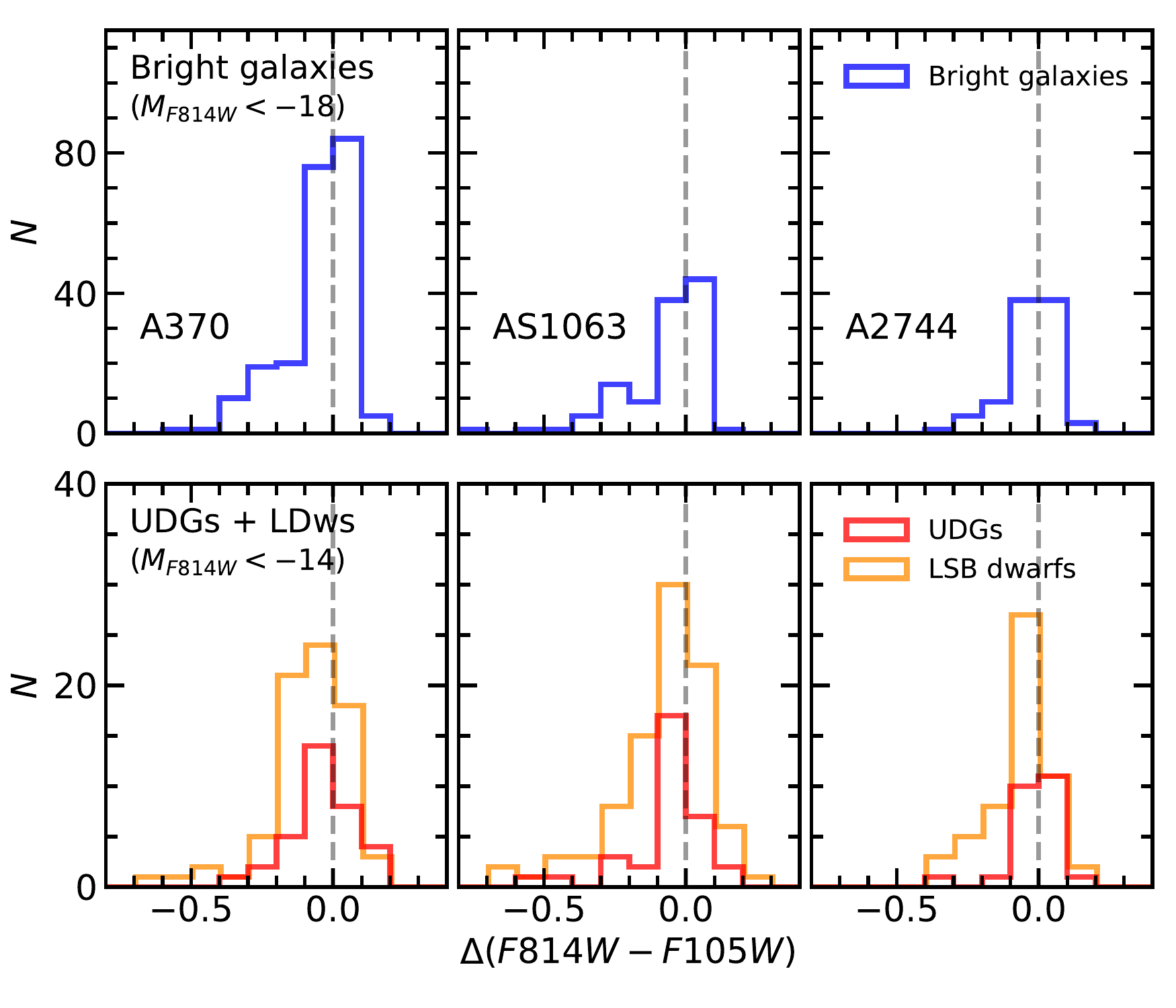}
	\caption{Histograms of color differences from the red sequences of the three HFF clusters (See {\color{blue}\bf {\bf Figure \ref{fig_cmd3}}}). Gray dashed lines denote the red sequence. The upper panels show the color distributions of bright galaxies (blue histograms), and the lower panels show those of UDGs (red histograms) and LSB dwarfs (yellow histograms). We select each galaxy population with the same absolute magnitude criteria: bright galaxies with $M_{F814W}<-18.0$ mag and LSB galaxies (UDGs and LSB dwarfs) with $-18.0<M_{F814W}<-14.0$ mag.
	\label{fig_chist}}
\end{figure}

\subsubsection{A Comparison of Abell 370 with Abell S1063 and Abell 2744}

In {\color{blue}\bf {\bf Figure \ref{fig_cmd3}}}, we display the CMDs of the central fields of Abell 370 (this work) and two other HFF clusters (Abell S1063 and Abell 2744) in \citet{Lee17}.
All these three HFF clusters show a prominent feature of the red sequence.
The linear fits of the sequence denoted by the black solid lines show a well-defined red sequence of each cluster.
The number of galaxies bluer than the red sequence in Abell 370 is larger than those in Abell S1063 and Abell 2744.
This implies that there are more galaxies in transition from blue star-forming galaxies to red quiescent galaxies in Abell 370 compared to the other two HFF clusters.

\begin{figure*}
	\centering
	\includegraphics[scale=0.55]{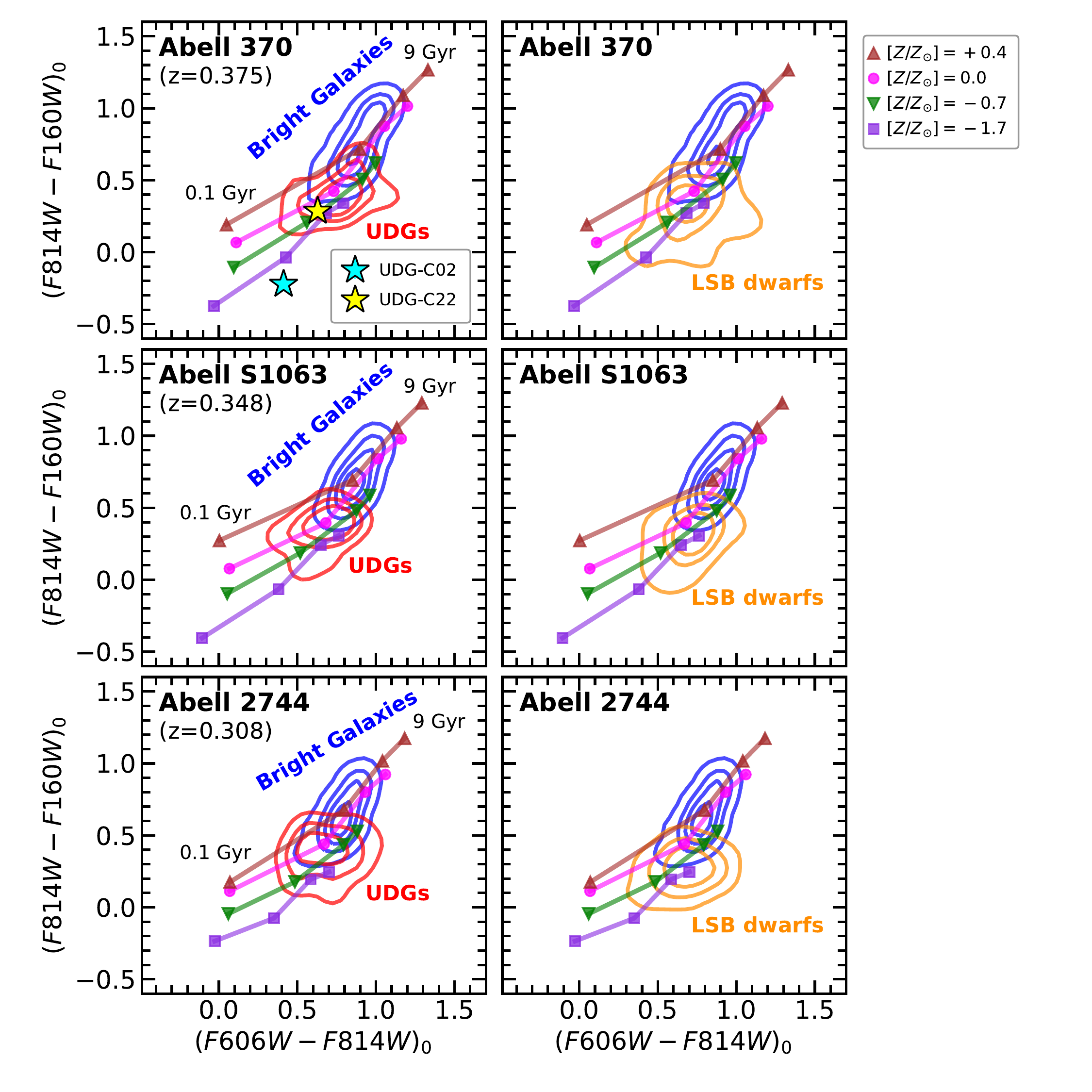}
	\caption{Color-color diagrams (CCDs) of bright galaxies, UDGs, and LSB dwarfs in the three HFF clusters. The left panels show the 2D color distributions of bright galaxies (blue contours) and UDGs (red contours) in each cluster. The right panels show the 2D color distributions of bright galaxies and LSB dwarfs (yellow contours). The SSP models have ages of 0.1, 1, 3, and 9 Gyr at the redshifts of the clusters and metallicities with $[Z/Z_{\odot}]=-1.7$ (purple squares), $-0.7$ (green upside-down triangles), $0.0$ (magenta circles), and $+0.4$ (brown triangles). In the top left panel, we mark UDG-C02 (cyan star) and UDG-C22 (yellow star).
	\label{fig_ccd}}
\end{figure*}

In {\color{blue}\bf {\bf Figure \ref{fig_chist}}}, we display the color distributions of each galaxy population in the central fields of the three HFF clusters.
For the x-axis, we calculate the color differences from the red sequences ($\Delta(F814W-F105W)$) of the three HFF clusters.
The upper panels show the color difference histograms of bright galaxies from the linear fits of the red sequences, and the lower panels show those of UDGs and LSB dwarfs.
The linear fits denoted by solid lines in {\color{blue}\bf {\bf Figure \ref{fig_cmd3}}} show the narrow red sequence very well.
The red sequence is plotted as gray dashed lines in {\color{blue}\bf {\bf Figure \ref{fig_chist}}}.
Overall, the three galaxy populations are dominantly located in the red sequence.
Bright galaxies show a bimodal color distribution with a large number of red galaxies and a small number of star-forming blue galaxies.
In the lower panels, LSB dwarfs seem to have a higher blue fraction ($f_{\rm blue}$) than UDGs.
Using blue galaxies with $\Delta(F814W-F105W)<-0.1$, we obtained $f_{\rm blue}=0.34\pm0.04~(76/223)$ for LSB dwarfs and $f_{\rm blue}=0.19\pm0.05~(17/91)$ for UDGs in the three HFF clusters.
Using blue galaxies with $\Delta(F814W-F105W)<-0.2$, we obtained $f_{\rm blue}=0.14\pm0.03~(31/223)$ for LSB dwarfs and $f_{\rm blue}=0.10\pm0.03~(9/91)$ for UDGs.
This indicates that UDGs could be older and more quiescent in their star formation than LSB dwarfs.
However, there are a small number of blue UDGs in each galaxy cluster, implying that star formation is not quenched in all UDGs.

\subsection{CCDs of the Galaxies}
In {\color{blue}\bf {\bf Figure \ref{fig_ccd}}}, we display the de-reddened CCDs ($(F606W-F814W)_{0}$ vs. $(F814W-F160W)_{0}$) of the galaxies in Abell 370 as well as those in Abell S1063 and Abell 2744.
We adopted the foreground extinctions of each filter as listed in \citet{shp18}.
We plot the number density contours of the bright galaxies in the red sequence, the UDGs, and the LSB dwarfs in these CCDs.
Then, we overlay the simple stellar population (SSP) models from GALAXEV. 
In order to construct the SSP models, we adopt the following options: 
spectral templates from \citet{bc03}, instantaneous burst star formation history (SFH), no dust, and the Chabrier initial mass function \citep{cha03}.
We set the range of ages of 0.1, 1, 3, and 9 Gyr at the redshifts of the three HFF clusters.
We set the range of metallicities with $[Z/Z_{\odot}]=-1.7$, $-0.7$, $0.0$ (solar), and $+0.4$.

The upper panels of {\color{blue}\bf {\bf Figure \ref{fig_ccd}}} show the de-reddened CCDs of bright red sequence galaxies, UDGs, and LSB dwarfs in Abell 370.
In this figure, the UDGs and the LSB dwarfs show a similar range of colors.
For Abell 370, the peak values of the color distribution of the UDGs ($(F606W-F814W)_{0}=0.79$ and $(F814W-F160W)_{0}=0.39$) and the LSB dwarfs ($(F606W-F814W)_{0}=0.65$ and $(F814W-F160W)_{0}=0.32$) are bluer than those of the bright red sequence galaxies ($(F606W-F814W)_{0}=0.88$ and $(F814W-F160W)_{0}=0.59$), as listed in {\color{blue}\bf {\bf Table \ref{tab_cc}}}.
These color ranges of the UDGs and the LSB dwarfs are consistent with the old-age SSP models for low metallicity ($[Z/Z_{\odot}]\lesssim-0.7$).
On the other hand, the colors of bright red sequence galaxies are better matched with the SSP models with higher metallicity ($[Z/Z_{\odot}]\gtrsim0.0$).
These trends of Abell 370 are similarly seen in the other HFF clusters, Abell S1063 and Abell 2744.

In the upper left panel of this figure, we mark UDG-C02 (the bluest UDG) and UDG-C22 (the largest UDG) of Abell 370.
UDG-C02 seems to have an age younger than the other UDGs by a few Gyr.
This implies that UDG-C02 includes recently formed stars.
In contrast, UDG-C22 shows red colors close to the mean values of the UDGs.
This means that UDG-C22 is made of old stars, which is similar to most UDGs in Abell 370.
These SSP models presume oversimplied SFHs and dust contents of galaxies, but they are still useful to compare the relative difference of ages and metallicities between bright galaxies and LSB galaxies.

\begin{figure}
	\centering
	\includegraphics[scale=0.55]{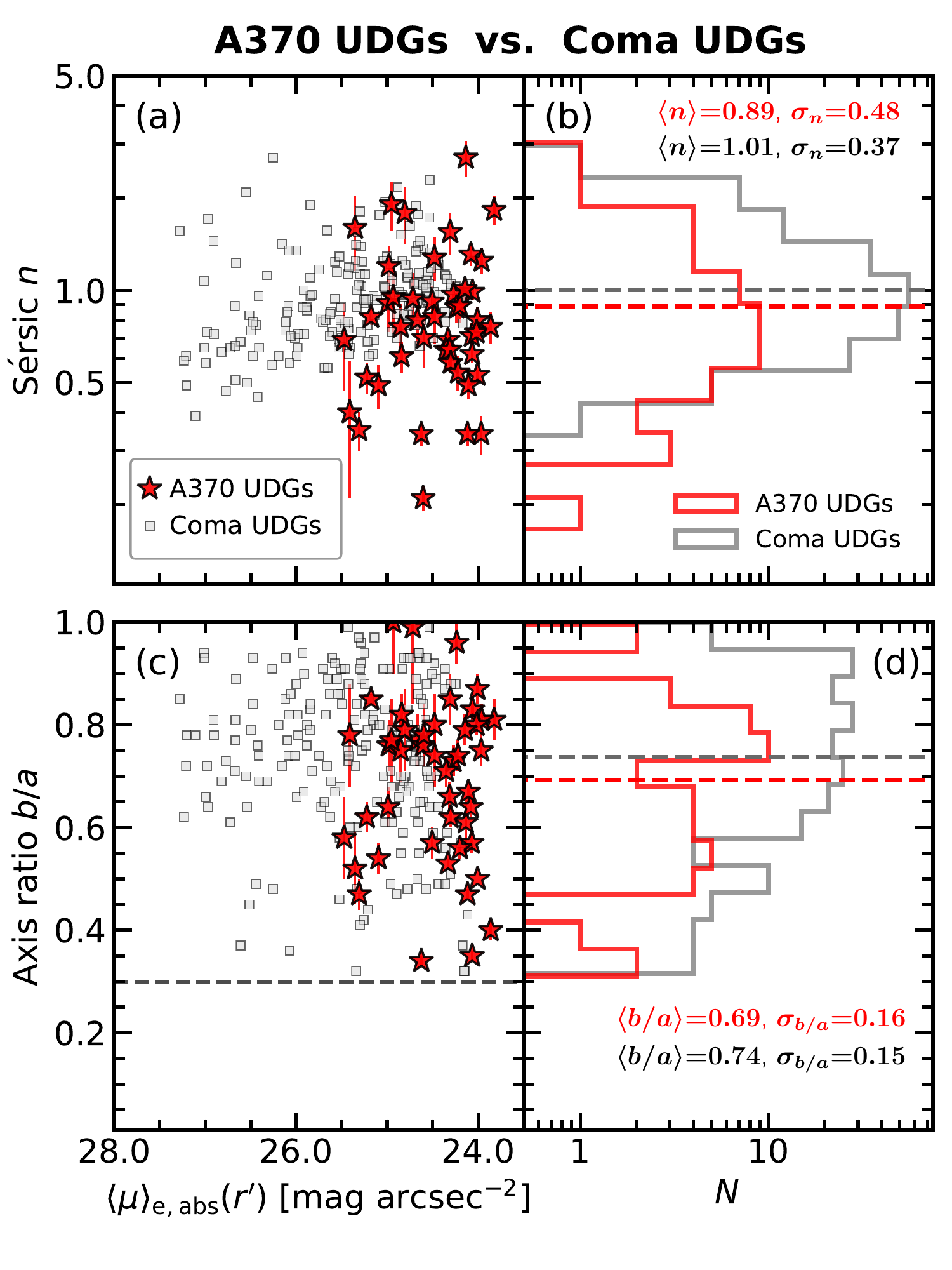}
	\caption{Comparison of structural parameters for UDGs in Abell 370 (red stars) and the Coma cluster (gray squares). The upper panels plot S\'ersic indices $n$ of Abell 370 UDGs and Coma UDGs as a function of \absmue$~$(a) and the corresponding histograms (b). The lower panels plot their axis ratios as a function of \absmue$~$(c) and the histograms (d). The mean and standard deviation values are marked in the right panels. In the lower left panel, gray dashed line denotes our selection criteria of axis ratios with $b/a>0.3$.
	\label{fig_str}}
\end{figure}

\subsection{Structural Parameters of the UDGs}

We estimate the structural parameters of the galaxies in Abell 370 such as S\'ersic indices ($n$), and axis ratios ($b/a$) using GALFIT. 
We compare these structural parameters of Abell 370 UDGs with those of the known UDGs in the Coma cluster.
The Coma cluster is known to host a large number of LSB galaxies \citep{yag16}, so that it serves as a very useful reference.
We select 193 Coma UDGs out of the whole sample of Coma LSBs given by \citet{yag16}, using the same size criteria as used in this work (\Reffc~$>1.5$ kpc).

{\color{blue}\bf {\bf Figure \ref{fig_str}}} shows the comparison of the structural parameters of Abell 370 UDGs with Coma UDGs.
The upper panels of {\color{blue}\bf {\bf Figure \ref{fig_str}}} compare the S\'ersic indices $n$ of UDGs in the two clusters.
Overall, the S\'ersic indices of UDGs in the two clusters have similar distributions with $\langle n\rangle\lesssim1$.
These results are consistent with those of UDGs and dwarf galaxies in other clusters such as Fornax and Abell 168 ($\langle n\rangle\sim0.7$) \citep{mun15, rom17a}.
In \citet{Lee17}, the mean values of the S\'ersic indices of UDGs in Abell S1063 and Abell 2744 are also $\langle n\rangle\simeq1.0$.
UDGs with $n>2$  are rare in these galaxy clusters. 
This indicates that most cluster UDGs have exponential light profiles regardless of their host cluster.

The lower panels of {\color{blue}\bf {\bf Figure \ref{fig_str}}} describe the distribution of axis ratios $b/a$ of UDGs in the two clusters.
In the Coma cluster, there are no UDGs with $b/a<0.3$, although \citet{yag16} did not use any selection criterion of $b/a>0.3$.
The axis ratios of UDGs in these two clusters also have consistent distributions with the mean values of $\langle b/a\rangle\sim0.7$.
These mean values are similar to those of UDGs in Abell S1063 ($\langle b/a\rangle=0.66$) and Abell 2744 ($\langle b/a\rangle=0.68$).
This means the morphology of cluster UDGs is closer to round shapes rather than to elongated shapes.

\begin{figure*}
	\centering
	\includegraphics[scale=0.7]{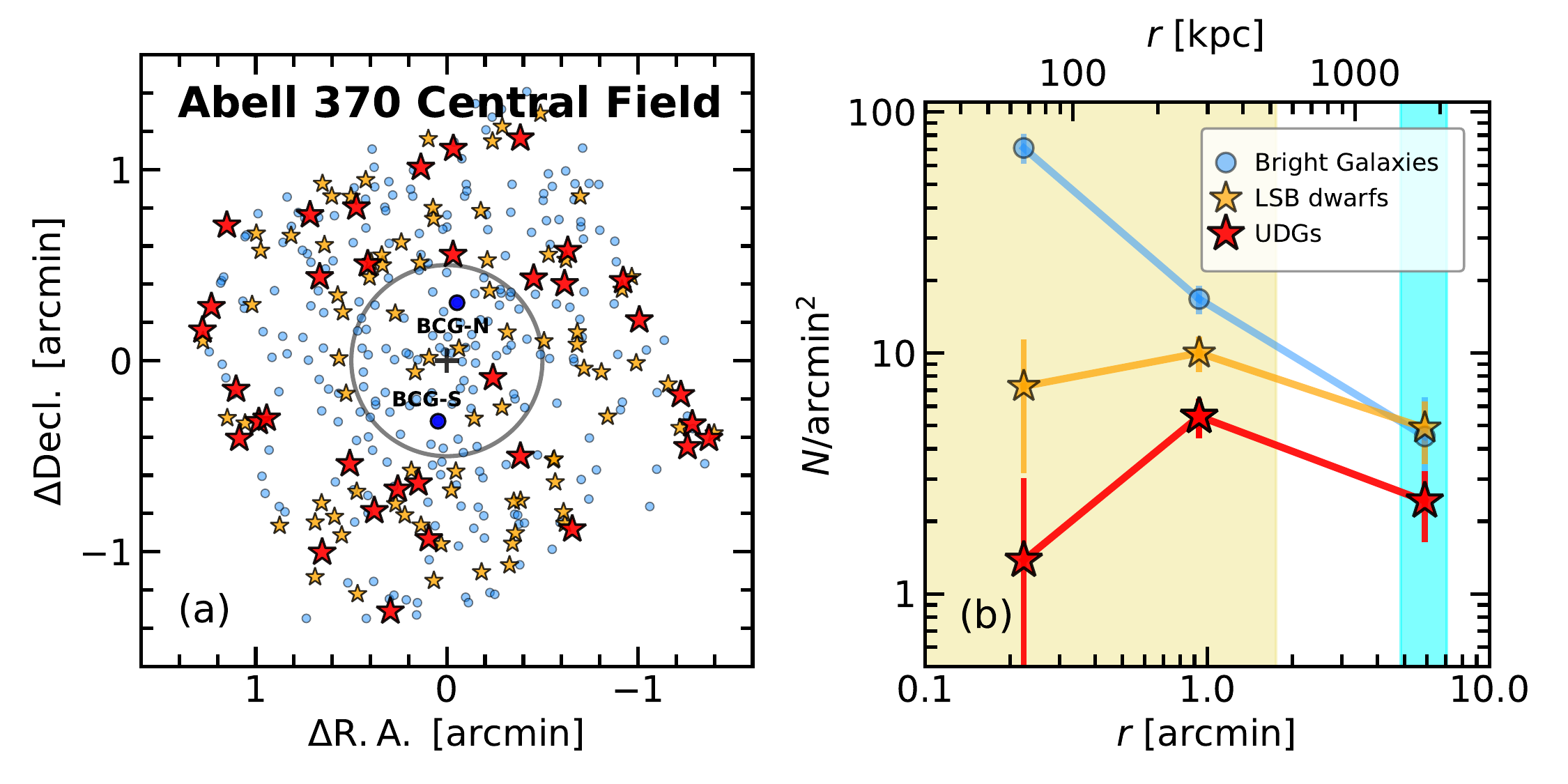}
	\caption{(a) Spatial distribution of galaxies in the central field of Abell 370. Symbols of galaxies are the same as in {\color{blue}\bf {\bf Figure \ref{fig_selection}}}, but here we plot only bright red sequence galaxies. Black crossmark denotes the center of Abell 370, and two large blue circles denote BCG-N and BCG-S. Gray circle with radius of $0\farcm5$ ($\sim$150 kpc) represents the boundary we used to divide the cluster central field. (b) Radial number density profiles (RDPs) of galaxies in the Abell 370 cluster. Blue circles, yellow stars, and red stars are the RDPs of the bright galaxies, LSB dwarfs, and UDGs. Yellow and cyan shaded regions represent the cluster central field and the parallel field.
	\label{fig_radial}}
\end{figure*}

\subsection{Spatial Distribution and RDPs of the UDGs}

In the left panel of {\color{blue}\bf {\bf Figure \ref{fig_radial}}}, we plot the spatial distributions of bright galaxies in the red sequence, UDGs, and LSB dwarfs in the central field of Abell 370.
The center of the Abell 370 cluster is set to be at the middle point between BCG-N and BCG-S (R.A. (2000)$=2^{\rm h}39^{\rm m}52\fs94$, Decl.(2000)$=-1^{\circ}34'37\farcs0$), as adopted by \citet{Lag17}.
We divide this field into two radial bins based on the clustercentric radius of $0\farcm5~(\sim150~\rm kpc)$ (the gray circle). 
Bright galaxies seem to be more centrally concentrated in the inner region ($r<0\farcm5$) than LSB galaxies.
It is noted that only one UDG is found in the inner region ($r<0\farcm5$), while most UDGs are seen in the outer region ($r>0\farcm5$).

We derive the RDPs of the bright red sequence galaxies, UDGs, and LSB dwarfs, using the data from the central field and the parallel field.
We estimate their background number density using the data for the XDF, and subtract them from the results of Abell 370.
We correct the resulting number density profiles using the completeness values derived from artificial galaxy tests (as described in \textbf{\S2.3} and {\color{blue}\bf {\bf Figure \ref{fig_mocktest2}}}).

The right panel of {\color{blue}\bf {\bf Figure \ref{fig_radial}}} shows the RDPs for Abell 370.
The RDP of the bright red sequence galaxies shows a clear central concentration.
In contrast, the central concentration of the UDGs and LSB dwarfs is much weaker than that of the bright red sequence galaxies.
It is noted that the RDPs of the UDGs and LSB dwarfs show a drop or flattening in the innermost bin ($r<0\farcm5$), while that of the bright red sequence galaxies keeps increasing as clustercentric distances decrease.

\begin{figure*}
	\centering
	\includegraphics[scale=0.55]{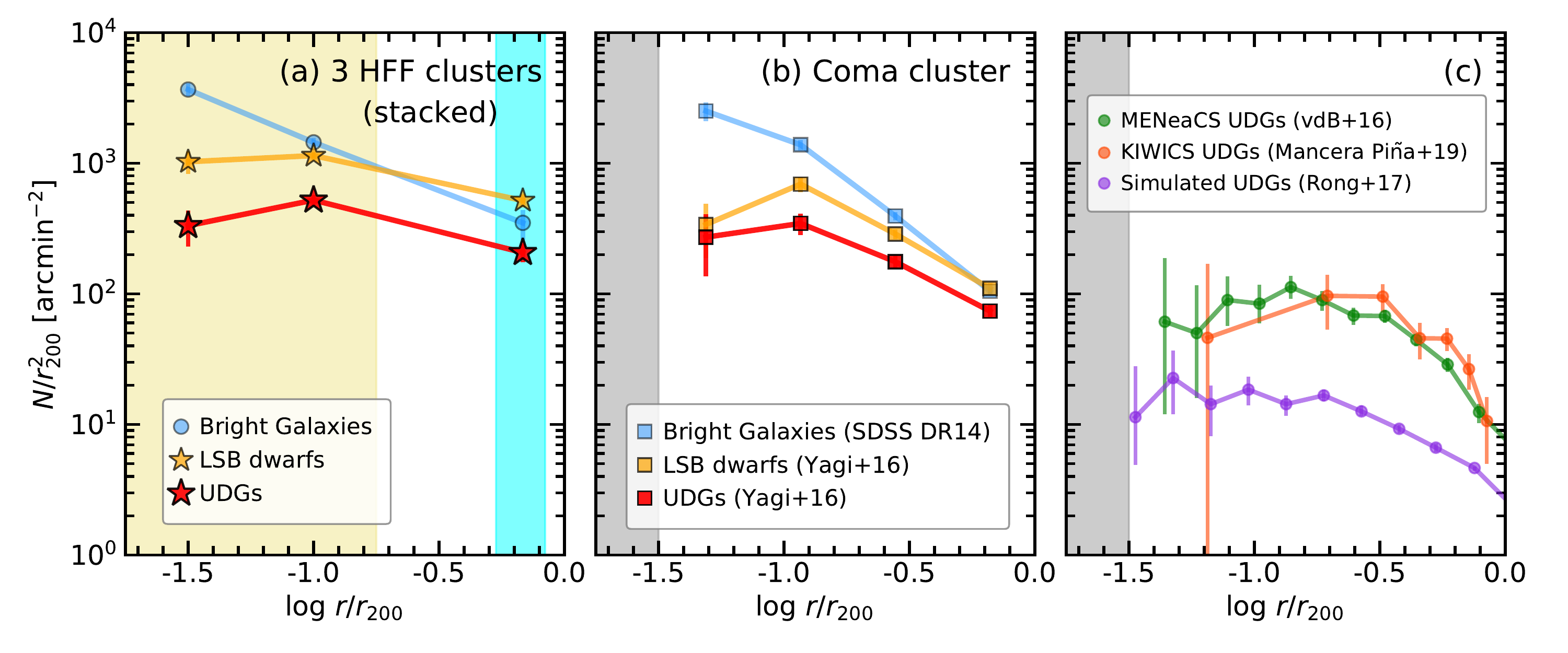}
	\caption{RDPs of cluster galaxies from observation and simulation results. (a) The stacked number density profiles for three HFF clusters (Abell 370, Abell S1063, and Abell 2744). Three galaxy populations are indicated: bright galaxies (blue circles), LSB dwarfs (yellow stars), and UDGs (red stars). (b) Same as (a), but for the Coma cluster. The bright galaxy samples (blue squares) are taken from SDSS DR14 (see texts for details). UDGs and LSB dwarfs are selected from \citet{yag16} with the same selection criteria as used in this work. (c) The RDPs of UDGs in eight MENeaCS clusters \citep{vdB16} (green circles), eight KIWICS clusters \citep{man19} (orange circles), and simulated clusters \citep{ron17} (purple circles).
	\label{fig_radialc}}
\end{figure*}

These features of RDPs for Abell 370 are similar to those of Abell S1063 and Abell 2744 (see Figure 7 in \citet{Lee17}). 
However, the numbers of UDGs and LSB dwarfs in the inner region of each cluster are too small to obtain statistically meaningful conclusions.
Therefore, we stack the RDPs of the three HFF clusters: Abell 370 (this work), Abell S1063, and Abell 2744. 
Before stacking we normalized the clustercentric distance with respect to the virial radius ($r_{200}$) of each cluster:
$r_{200}=8\farcm52$ for Abell 370, $8\farcm64$ for Abell S1063, and $9\farcm16$ for Abell 2744. 
In {\color{blue}\bf {\bf Figure \ref{fig_radialc}}} (left panel), we display the stacked RDPs of the bright red sequence galaxies, UDGs, and LSB dwarfs.
Again, the stacked RDPs of the UDGs and LSB dwarfs show a flattening or drop in the inner region (log $r/r_{200}<-1.0$), in contrast to the RDP of the bright red sequence galaxies which rises as clustercentric distances decrease.
\citet{jan19} also found similar central depletion or flattening of the RDPs of UDGs in all the six HFF clusters.

In addition, we compare the RDPs of the HFF clusters with those of other clusters.
The middle panel of {\color{blue}\bf {\bf Figure \ref{fig_radialc}}} shows the RDPs of the three galaxy populations in the Coma cluster.
We select bright red sequence galaxies in the Coma cluster from SDSS DR14 \citep{abo18} by using the following criteria: magnitude ($r'<18$ mag), color (($g'-r'$) color for the red sequence of the Coma cluster), clustercentric distance ($r_{\rm cl}<r_{200}= 97\farcm92$ \citep{kub07}), and radial velocity ($v_{r}({\rm Coma})-v_{r}(r_{\rm cl})<v_{r}<v_{r}({\rm Coma})+v_{r}(r_{\rm cl})$), where $v_{r}({\rm Coma})$ is the systematic velocity of the Coma cluster ($v_{r}({\rm Coma})=6925.0~{\rm km~s^{-1}}$ in \citet{str99}) and $v_{r}(r_{\rm cl})=(GM_{200}r_{\rm cl}^{-1})^{1/2}$.
We divide Coma LSB galaxies in \citet{yag16} into UDGs ($R_{\rm eff,c} > 1.5$ \kpc) and LSB dwarfs ($R_{\rm eff,c} = 1.0-1.5$ \kpc), applying the same criteria as in this study.
The RDPs of the bright red sequence galaxies, UDGs, and LSBs in Coma show very similar trends to those of the three HFF clusters.

In the right panel, we display the RDPs of UDGs in nearby low-mass clusters in the literature: 8 MENeaCS clusters ($z=0.044-0.063$) with median virial mass of $M_{200}=5.55\times10^{14}$ $M_{\odot}$ in \citet{vdB16}, and
8 KIWICS clusters ($z=0.02-0.03$) with median virial mass of $M_{200}=3.35\times10^{13}~M_{\odot}$ in \citet{man19}. 
The RDPs of UDGs in these low-mass clusters show a trend consistent with those of the HFF clusters and the Coma cluster.

In the same figure, we overlay the RDP of the UDGs in simulated clusters given by \citet{ron17}.
These simulated clusters also have low masses of $M_{200}=10^{13}-10^{14}~M_{\odot}$.
The RDP of the simulated UDGs similarly shows a flattening in the inner region.
The flattening or dropping features of the RDPs of UDGs appear to be universal in galaxy clusters regardless of virial masses or redshifts of their host clusters.
From the comparison of the RDPs of UDGs from low to high mass clusters and in simulated clusters, we conclude that the RDPs of UDGs (as well as LSB dwarfs) show a flattening in the inner region of the clusters.
We discuss this aspect further in \textbf{\S4.1}.

\begin{figure}
	\centering
	\includegraphics[scale=0.67]{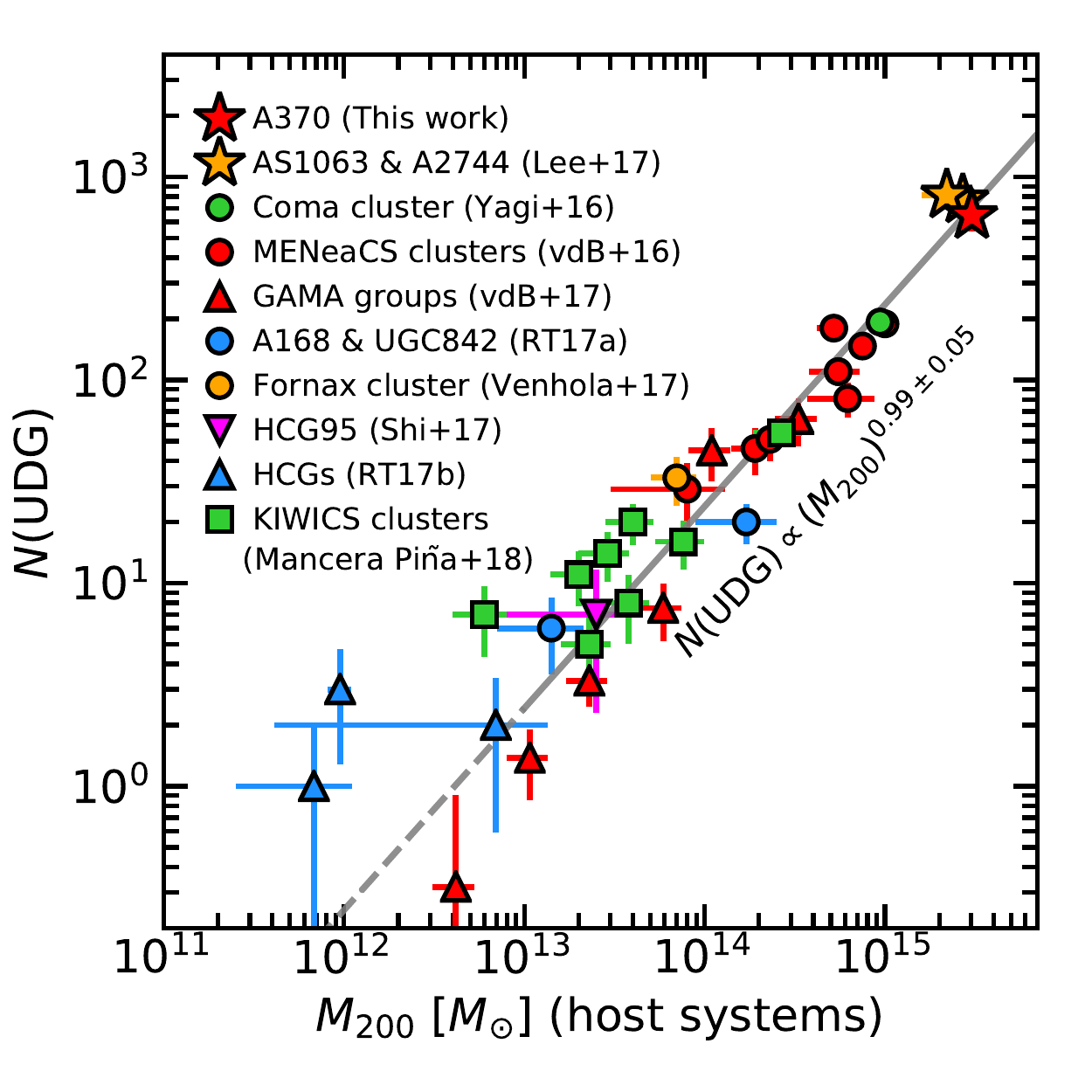}
	\caption{The UDG abundance versus virial mass of their host system ($M_{200}$) for Abell 370 (red star) in comparison with other systems in the previous studies. Gray solid line shows a power-law fit for the UDGs in the host systems with $M_{200}>10^{13}$ $M_{\odot}$: ${\rm log}~N({\rm UDG})=(0.99\pm0.05)\times{\rm log}~M_{200}+(-12.53\pm0.67)$.
	\label{fig_mn}}
\end{figure}

\subsection{The Abundance of UDGs and the Virial Masses of their Host Systems}
The abundance (total number) of galaxies inside the virial radii ($r_{200}$) of their host systems can help us understand the correlation between the number of galaxies and their environments. 
The abundance of galaxies 
UDGs ($N$(UDG)) is known to have a power-law relation with the virial masses ($M_{200}$) of their host systems: $N({\rm UDG})\propto M_{200}^{\alpha}$ \citep{vdB16, Lee17, rom17b, vdB17, man18}.
In this relation, the $\alpha$-value is a key parameter to determine how efficiently galaxies are formed and survive in their environments.
If $\alpha<1$, galaxies in low density environments have relatively higher number densities per mass of their host systems.
This implies that the galaxies are preferentially formed and survive in fields or galaxy groups rather than in massive galaxy clusters. 
In contrast, $\alpha>1$ means that galaxies are formed more efficiently or survive longer in high density environments.
$\alpha=1$ means that the number of galaxies simply depends on the masses of their host systems.
These galaxies do not strongly depend on environmental effects.

We estimate the abundance of UDGs ($N$(UDG)) within the virial radius of Abell 370, by integrating the RDPs from the inner region to the outskirts at its virial radius.
We obtained a value of $N$(UDG)$=644\pm104$.
This value is consistent, within the error, with the value given by \citet{jan19}, $N$(UDG)$=711^{+213}_{-210}$ for Abell 370.

We counted UDGs in other host systems from previous studies, using the same criterion as used in this study.
{\color{blue}\bf {\bf Table \ref{tab_ab}}} provides a list of $N$(UDG) and $M_{200}$ of the host systems.
In {\color{blue}\bf {\bf Figure \ref{fig_mn}}}, we show the relation between $N$(UDG) and $M_{200}$ of Abell 370, in comparison with those of UDGs in other host systems in the literature: massive galaxy clusters \citep{yag16, Lee17}, low-mass clusters \citep{vdB16, rom17a, ven17, man19}, galaxy groups \citep{vdB17}, and compact galaxy groups \citep{rom17b, shi17}.

In the figure, $N$(UDG) and $M_{200}$ show a tight correlation, but the scatter increases at low-mass host systems.
This scatter is significantly large for the mass of $M_{200}<10^{13}$ $M_{\odot}$, because low-mass host systems have only a small number of UDG.
Fitting the data for  $M_{200}>10^{13}M_{\odot}$,
we obtain ${\rm log}~N({\rm UDG})=(0.99\pm0.05)\times{\rm log}~M_{200}+(-12.53\pm0.67)$ with a root mean square (RMS) of 0.19 dex, where the value of $\alpha$ is basically one.
Similarly, we derive $\alpha=0.97\pm0.05$ for $M_{200}>10^{12}M_{\odot}$.
If we use the entire range of the virial mass for fitting,
we obtain $\alpha=0.92\pm0.05$.
Thus, the $\alpha$-value is close to one.
This indicates that the efficiency of the formation or survival of UDGs is little dependent on their environments.

If we fit the virial mass in terms of UDG abundance for $M_{200}>10^{13}M_{\odot}$,
we obtain ${\rm log}~M_{200}=(1.01\pm0.05)\times{\rm log}~N({\rm UDG})+(12.61\pm0.10)$ with an RMS value of $0.19$ dex.
This relation can be used to estimate the virial masses of the host systems using the abundance of UDGs.

\subsection{Dynamical Mass of the UDGs}
Dynamical mass of UDGs is a critical parameter to understand the nature of UDGs.
Various methods have been applied to estimate dynamical masses of UDGs in the literature.
First, direct measurements of velocity dispersions of UDGs with spectroscopy can be used to estimate their virial masses under the assumption that UDGs are pressure-supported systems.
For example, \citet{mar19} obtained $\sigma_{v}=56^{+10}_{-10}$ \kms, and $M_{200}\sim5\times10^{11}~M_{\odot}$ for DGSAT-I.
Second, the total number of GCs is useful for mass estimation of their hosts.
There are well-known relations between the number of GCs ($N_{\rm GC}$) and halo masses ($M_{\rm halo}$) (\citet{har17} and references therein).
Since this method is easier than spectroscopy, a number of studies applied this method to the sample of Coma UDGs including DF17 ($N_{\rm GC}=25\pm11$ and $M_{\rm halo}\sim9\times10^{10}~M_{\odot}$), DF44 ($N_{\rm GC}=76\pm18$ and $M_{\rm halo}\sim8\times10^{11}~M_{\odot}$), and DFX1 ($N_{\rm GC}=63\pm17$ and $M_{\rm halo}\sim5\times10^{11}~M_{\odot}$) \citep{bea16b, pen16, van16, van17, amo18b, Lim18}. 
Third, HI line widths can be used to estimate the dynamical mass of gas-rich UDGs. 
\citet{tru17} measured the HI line width of UGC 2162, a blue isolated gas-rich UDG, to be $W(\rm HI)=126$ \kms~ and derived a virial mass of $M_{200}\sim8\times10^{10}$ $M_{\odot}$.
Fourth, weak lensing analysis can be used to estimate virial mass of UDGs.
\citet{sif18} used this method and constrained the maximum virial mass range of UDGs in the MENeaCS clusters: $M_{200}<6.3\times10^{11}~M_{\odot}$ with the 95\% confidence level.

In the case of Abell 370 UDGs/LSB dwarfs, none of the above methods can be used.
We estimate dynamical masses of UDGs/LSB dwarfs approximately, using the fundamental manifold method based on photometric parameters as suggested by \citet{zar17}.
\citet{zar08} suggested a fundamental manifold of galaxies, with which we can derive kinematic terms from their sizes and surface brightness.
The kinetic term is defined by $V=\sqrt{\sigma_{v}^{2}+v_{\rm rot}^{2}/2}$, where $\sigma_{v}$ is the velocity dispersion and $v_{\rm rot}$ is the rotation velocity.
\citet{zar08} used 1,925 galaxies from bright ellipticals ($-22.0<M_{r'}<-18.5$) to low-mass disk galaxies ($-16.0<M_{r'}<-13.5$) \citep{ge06, zar06a, zar06b, piz07, sige07, spr07} in order to empirically calibrate the scaling relations.
They suggested that the derived relations can be applied to any types of galaxies regardless of their morphology.
\citet{zar17} applied this method to estimate the kinetic terms of LSB galaxies including UDGs after performing an observational correction (see the equations (1), (2) in \citet{zar17}).
Then, they estimated the virial masses of the UDGs using these derived kinetic terms.
\citet{Lee17} and \citet{cha18} also used this method to estimate dynamical masses of UDGs in massive clusters and simulations.
We estimate dynamical masses of UDGs and LSB galaxies in Abell 370, assuming that these galaxies are pressure-supported systems and their halo mass profiles follow the NFW profiles (see the equation (3) in \citet{Lee17} and the related text).

\begin{figure}
	\centering 
	\includegraphics[scale=0.45]{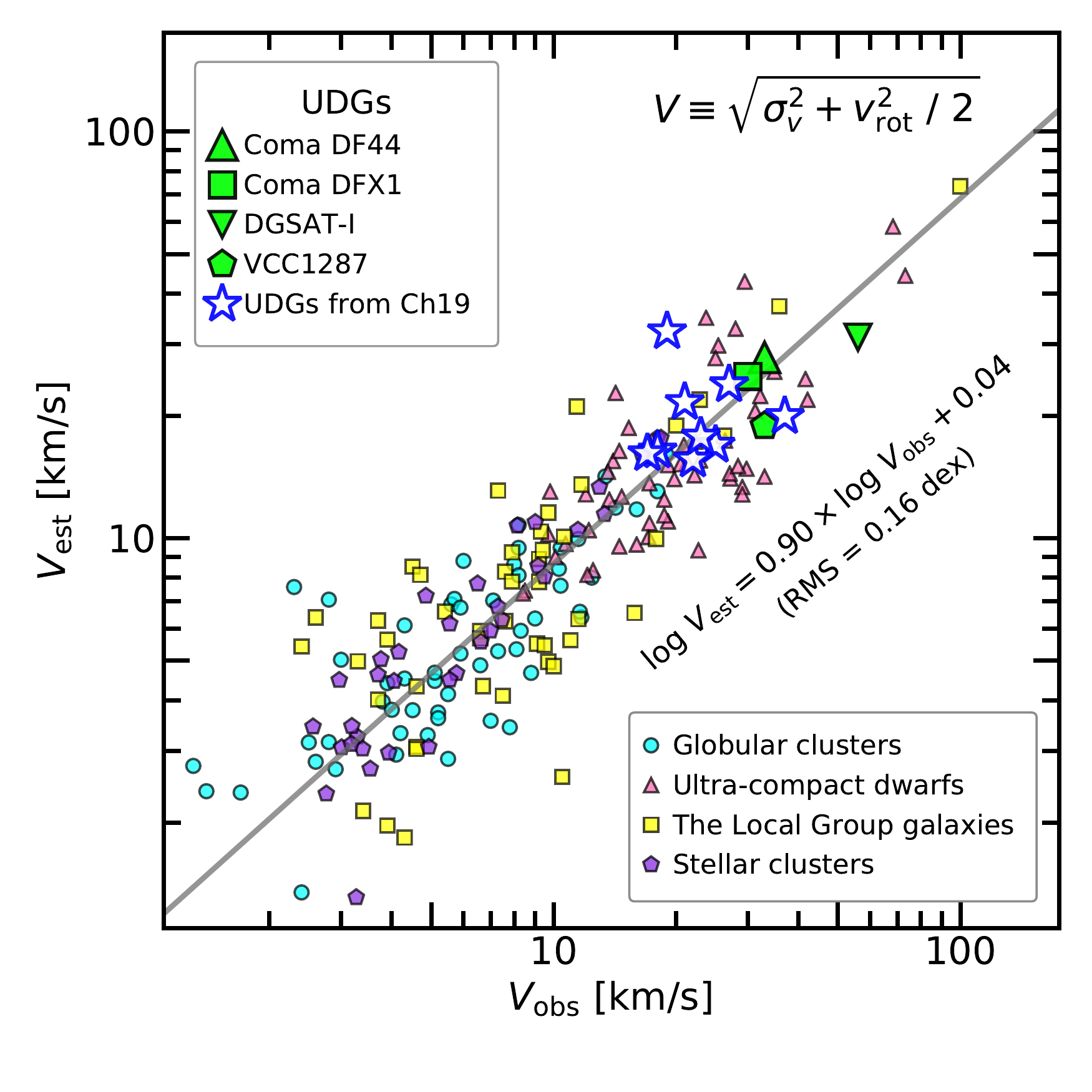}
	\caption{Comparison of the kinetic terms 
	($V\equiv\sqrt{\sigma_v^2 + v_{\rm rot}^2/2}$, $V=\sigma_v$ for $v_{\rm rot}=0$) of passive stellar systems derived from the spectroscopic measurement ($V_{\rm obs}$, x-axis) and the fundamental manifolds ($V_{\rm est}$, y-axis).
	We plot the data of 13 UDGs/LSB dwarfs with spectroscopic measurements of their velocity dispersions: Coma DF44 \citep{van19}, Coma DFX1 \citep{van17}, DGSAT-I \citep{mar19}, VCC 1287 \citep{bea16a}, and 9 Coma UDGs from \citet{chi19} (noted as `Ch19' in the figure).
	For comparison, we  added the sample of globular clusters \citep{Mc05}, ultra-compact dwarfs (UCDs) \citep{mie08, chi11, set14, ahn17, ahn18, afa18}, satellite galaxies in the Local Group \citep{Mc12}, and stellar clusters \citep{zar12, zar13, zar14}.
	The gray solid line shows the power-law fitting result of the relation between $V_{\rm obs}$ and $V_{\rm est}$: ${\rm log}~V_{\rm est}=(0.90\pm0.03)\times{\rm log}~V_{\rm obs}+(0.04\pm0.03)$ with an RMS value of 0.16 dex.
	\label{fig_z17}}
\end{figure}

However, \citet{zar17} used only two UDGs with spectroscopic measurements
(DF44 and VCC 1287) to check the validity of applying the fundamental manifold method to UDGs.  
Here we check it again using an increased sample of 13 UDGs/LSB dwarfs which have velocity dispersion measurements: 9 Coma UDGs in \citet{chi19}, Coma DF44 \citep{van19}, Coma DFX1 \citep{van17},  DGSAT-1 \citep{mar19}, and VCC 1287 \citep{bea16a}. 
6 of the 9 Coma UDGs in \citet{chi19} have effective radii smaller than 1.5 kpc, so they correspond to LSB dwarfs in this study. 

In {\color{blue}\bf Figure \ref{fig_z17}}, we plotted estimated kinetic terms ($V_{\rm est}$) versus observed kinetic terms ($V_{\rm obs}$) for these 13 UDGs/LSB dwarfs (it is an updated version of  Fig. 1 in \citet{zar17}).
For comparison, we plotted also the sample of GCs \citep{Mc05}, ultra-compact dwarfs (UCDs) \citep{mie08, chi11, set14, ahn17, afa18, ahn18}, satellite galaxies in the Local Group \citep{Mc12}, and stellar clusters \citep{zar12, zar13, zar14}.
We assumed $v_{\rm rot}=0$ except for a few dwarf galaxies in the Local Group and UCDs.
This figure shows that the 13 UDGs/LSB dwarfs follow very well the relation of other passive stellar systems.
We fit the data for all stellar systems including UDGs/LSB dwarfs with a power law, obtaining ${\rm log}~V_{\rm est}=(0.90\pm0.03)\times{\rm log}~V_{\rm obs}+(0.04\pm0.03)$ with an RMS value of 0.16 dex. 

This relation is very similar to the relation for the non-UDG sample given by \citet{zar17},
${\rm log}~V_{\rm est}=0.88\times{\rm log}~V_{\rm obs} + 0.07$.
This indicates that the fundamental manifold method can be applied to the UDG/LSB dwarf regime.

\begin{figure}
	\centering 
	\includegraphics[scale=0.45]{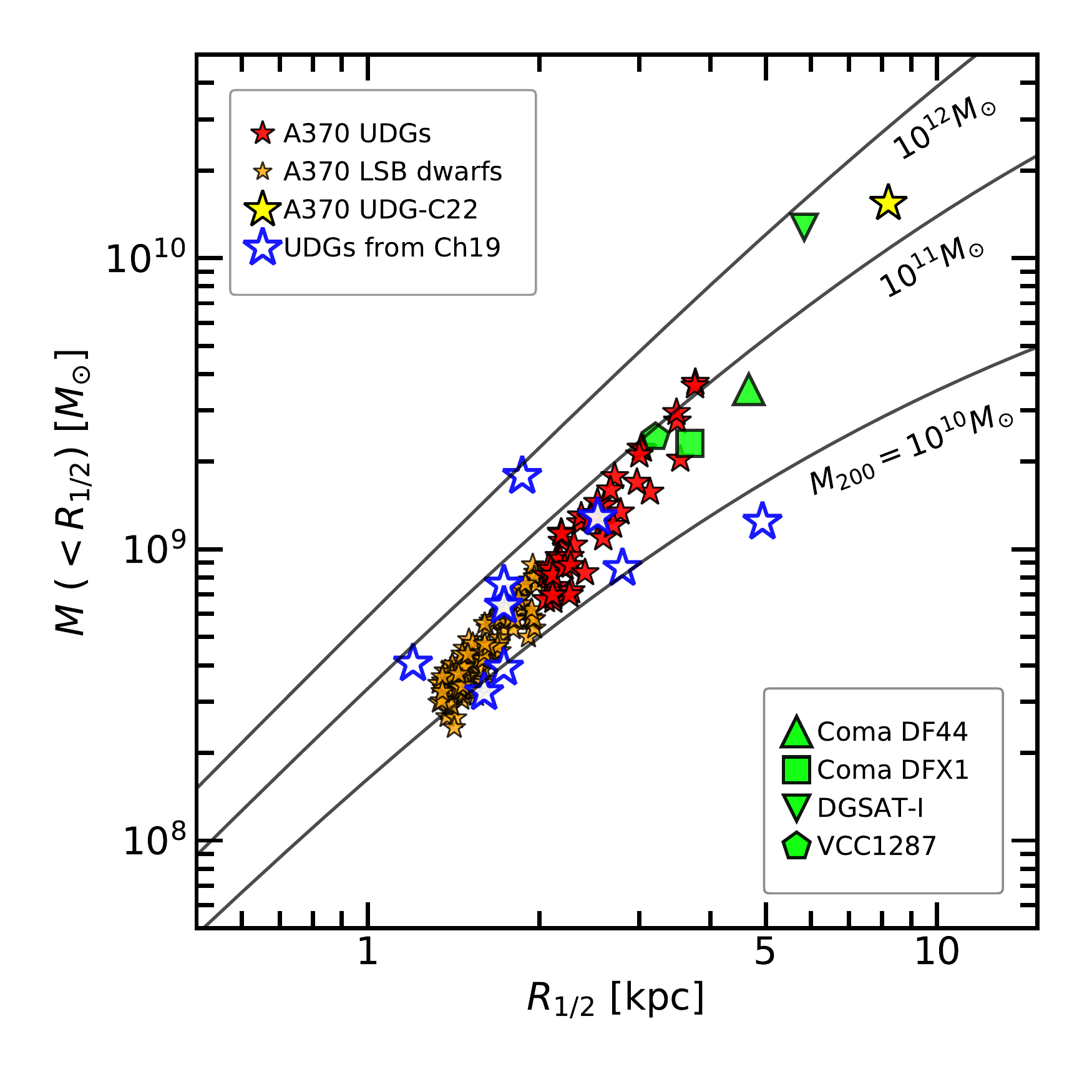}
	\caption{
	Enclosed masses ($M(<R_{1/2})$) vs. 3D half-light radii 
	($R_{1/2}= 4/3$ \Reffc) for Abell 370 UDGs (red star symbols) and LSB dwarfs (orange star symbols) in comparison with other galaxies in the literature (green and blue symbols as in Figure \ref{fig_z17}).
	Enclosed masses are derived from the scaling relations in \citet{zar08} and \citet{zar17}. 
	Black solid line curves 
	denote virial masses of $M_{200}$= $10^{10}$, $10^{11}$, and $10^{12}$ $M_{\odot}$ derived 
	from the NFW mass density profiles. 
	The large cyan star symbol is Abell 370 UDG-C22, the largest one.
	\label{fig_vmass}}
\end{figure}

{\color{blue}\bf {\bf Figure \ref{fig_vmass}}} displays the dynamical mass distributions of UDGs and LSB dwarfs in Abell 370 derived from the fundamental manifold method in this study.
We marked UDG-C22, the largest one, by a yellow star symbol.
The x-axis is the 3D half-light radius ($R_{1/2}=4R_{\rm eff,c}/3$), and the y-axis is the enclosed dynamical mass within $R_{1/2}$.
The curved solid lines denote the virial masses ($M_{200}$) of $10^{10}$, $10^{11}$, and $10^{12}~M_{\odot}$ derived from the NFW mass profiles.
For comparison, we also plotted other known UDGs/LSB dwarfs with spectroscopic measurements in the literature \citep{bea16a, van16, van17, chi19, mar19} as in {\color{blue}\bf Figure \ref{fig_z17}}.

Several features are noted in {\color{blue}\bf {\bf Figure \ref{fig_vmass}}}.
First, the locations of Abell 370 UDGs/LSB dwarfs are overlapped  with those of other UDGs with spectroscopic measurements. This implies that the dynamical mass range of Abell 370 UDGs/LSB dwarfs is similar to that of the other UDGs/LSB dwarfs.
Second, UDGs have relatively larger enclosed masses than LSB dwarfs. Thus, UDGs are larger and more massive than LSB dwarfs.
Third, most UDGs and LSB dwarfs have virial masses with $M_{200}=10^{10}-10^{11}$ $M_{\odot}$.
This implies that a majority of UDGs are dwarf-like galaxies.
Fourth, 
a few UDGs in Abell 370 have larger virial masses, $10^{11}<M_{200}<10^{12}$ $M_{\odot}$.
Among them, UDG-C22, the largest one in Abell 370, has the largest mass. 
Interestingly DGSAT-I, the largest in the other UDG sample, has a mass similar  to that of UDG-C22.

\begin{figure*}
	\centering
	\includegraphics[scale=1.0]{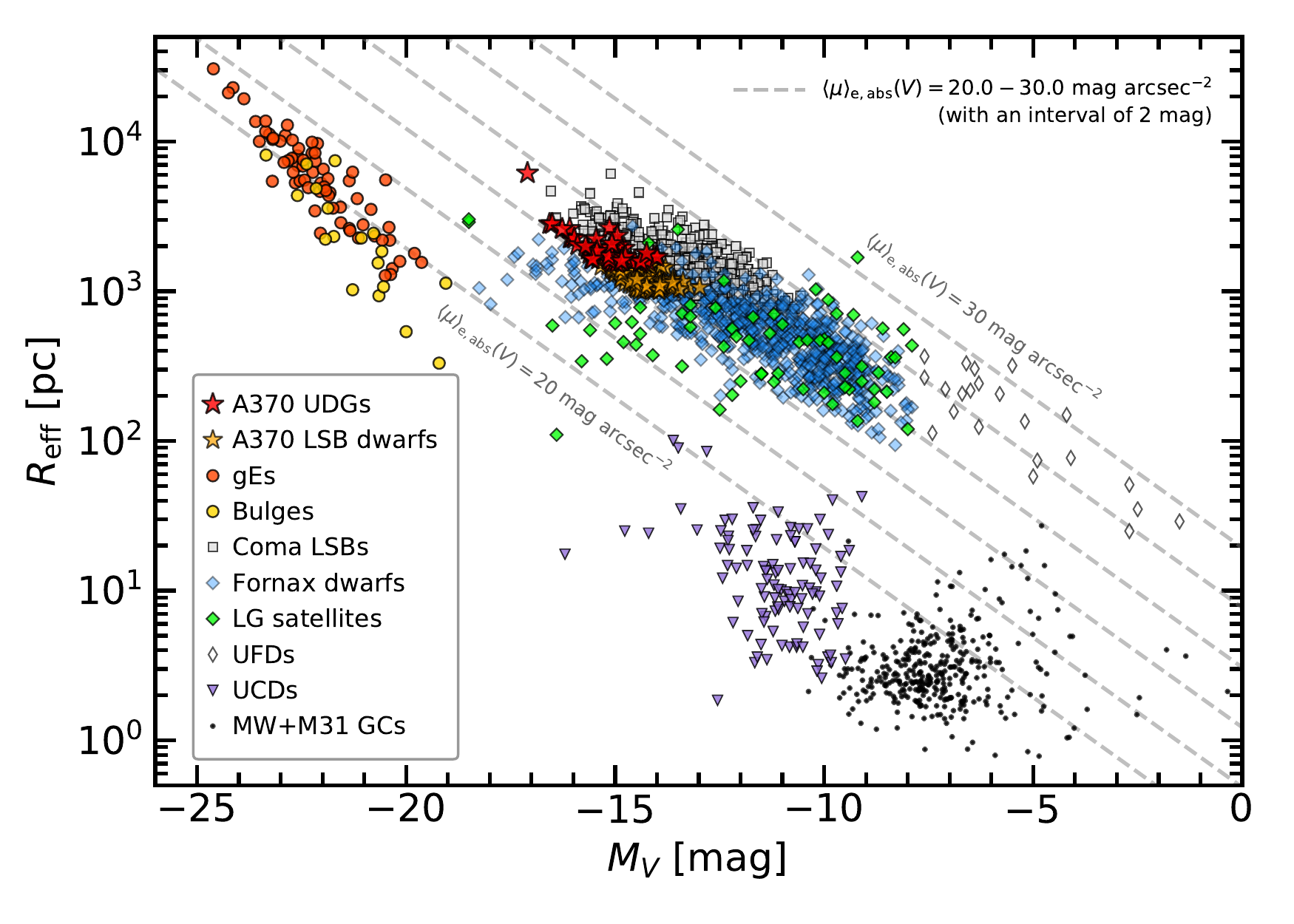}
	\caption{The size-luminosity relation diagram for Abell 370 UDGs (red star symbols) and LSB dwarfs (orange star symbols) in comparison with other passive stellar systems:
	giant ellipticals (gEs) and bulges in spirals (red and yellow circles) \citep{ben93}, Coma LSBs (gray squares) \citep{yag16}, Fornax dwarfs (blue diamonds) \citep{mun15, ord18}, the Local Group satellites (green diagmonds) \citep{Mc12}, ultra-compact dwarfs (purple upside-down triangles) \citep{pan16}, and GCs in the MW and M31 (black dots) \citep{har10, pea10}.
	Ultra-faint dwarfs (UFDs, white diamonds) are selected with $M_{V}>-7.7$ from the sample of dwarf galaxies.
	Gray dashed lines denote the surface brightness of $\langle\mu\rangle_{\rm e,abs}(V)=20$, $22$, $24$, $26$, $28$, and $30~{\rm mag~arcsec^{-2}}$ from left to right.
	\label{fig_scaling}}
\end{figure*}

\subsection{The Size-Luminosity Relation of UDGs and LSB Dwarfs}

We display the size-luminosity relation of  UDGs and LSB dwarfs of Abell 370 in {\color{blue}\bf {\bf Figure \ref{fig_scaling}}}.
For comparison, we also plotted other passive stellar systems: giant ellipticals and bulges in spirals \citep{ben93}, Coma LSBs \citep{yag16}, Fornax dwarfs \citep{mun15,ord18}, UCDs \citep{pan16}, dwarf galaxies in the Local Group \citep{Mc12}, and GCs in the Milky Way (MW) and M31 \citep{har10,pea10}.
Ultra-faint dwarfs (UFDs, $M_V>-7.7$ mag) in the Local Group are marked by thin diamonds.
In the figure, we added the loci of
surface brightness for $\langle\mu\rangle_{\rm e,abs}(V)=20- 30~{\rm mag~arcsec^{-2}}$ with an interval of 2 magnitude.

{\color{blue}\bf {\bf Figure \ref{fig_scaling}}} shows following features.
First, UDGs/LSB dwarfs in Abell 370 occupy the larger and brighter side of UDGs/LSB dwarfs in Coma, Fornax, and the Local Group.
Second, UDGs in Abell 370 are just a larger and brighter version of LSB dwarfs, not showing any clear distinction between the two types.
Third, UDGs/LSB dwarfs are obviously separated from the high surface bright objects such as giant ellitical galaxies and spiral bulges, UCDs, and globular clusters.

\section{Discussion}

\subsection{Implications of RDPs of the UDGs}
In \textbf{\S3.4}, we find that the RDPs of UDGs show a flattening (or a slight drop) in the inner region of their host systems, while the RDPs of bright red sequence galaxies do not.
This can be seen from low-mass clusters to high mass clusters.
The RDPs of UDGs and LSB dwarfs in the three HFF clusters also show a similar trend.

Most previous studies interpreted that tidal disruptions of UDGs contributed to their low number densities in the central region.
\citet{vdB16} and \citet{Lee17} suggested that some UDGs were disrupted by the strong gravitational potential near the center of clusters.
They noted that most UDGs could be easily disrupted because of their low masses.
In addition, \citet{ven17} and \citet{wit17} also found that the number densities of UDGs in the central region of the Fornax cluster and the Perseus cluster are lower than in the outer region.
These studies pointed out that it is difficult for UDGs to survive in the strong tidal interactions of the cluster central regions.

Observationally, tidally disrupted UDGs are frequently found from the local Universe to distant galaxy clusters.
UDGs around massive galaxies such as And XIX (near M31) \citep{col13, col19}, CenA-MM-Dw3 (near NGC 5128) \citep{crn16}, and Scl-MM-Dw2 (near NGC 253) \citep{tol16} show tidal features toward their neighboring massive galaxies. 
In cluster environments, several UDGs in the Virgo cluster \citep{mih15, tol18} and the HFF clusters \citep{Lee17} also have tidally disrupted features near bright galaxies in the clusters.
Thus, our results support the tidal disruption scenario for low number densities of UDGs as well as LSB dwarfs in the central regions of their host systems.

\subsection{Comparison of the Relation between UDG Abundance and Virial Mass of Their Host Systems}
In this study, we find that the relation between UDG abundance and the virial mass of their host systems is described well by a simple power-law relation with an index of $\alpha=0.99\pm0.05$ for $M_{200}>10^{13}~M_{\odot}$, and
$\alpha=0.97\pm0.05$ for $M_{200}>10^{12}~M_{\odot}$.
If we include two Hickson compact groups (HCGs)  with lower mass ($5\times 10^{11}~M_{\odot}<M_{200}<10^{12}~M_{\odot}$), we derive a slightly smaller value, $\alpha=0.92\pm0.05$.
It is noted that we selected UDGs in the literature samples using the same selection criteria as used in this study.

We compare our results with those in the previous studies, as shown in {\color{blue}\bf Figure \ref{fig_mn}} in the following.
\citet{vdB16} suggested $\alpha=0.93\pm0.16$ from UDGs in eight MENeaCS galaxy clusters at $z=0.044-0.063$.
Later, adding the stacked numbers of UDGs in galaxy groups from KiDS+GAMA fields, \citet{vdB17} derived $\alpha=1.11\pm0.07$.
\citet{Lee17} added 
the data for UDGs in two massive clusters (Abell S1063 and Abell 2744) to the previous sample, and obtained $\alpha=1.05\pm0.09$ 
for the systems with $M_{200}>10^{13}M_{\odot}$.
Recently, \citet{man18} suggested $\alpha=0.96\pm0.11$ for $M_{200}=10^{12}- 5\times 10^{14}~M_{\odot}$, using the samples in 8 clusters from the KIWICS and applying homogeneous selection criteria of UDGs.
These values are all close to one, being consistent with the results in this study.

On the other hand, a few studies presented slightly lower values of $\alpha$. 
\citet{rom17b} added UDGs in HCGs ($5\times 10^{11} M_{\odot} < M_{200}<10^{13}M_{\odot}$) to the sample and derived $\alpha=0.85\pm0.05$.
\citet{man18} also presented $\alpha=0.77\pm0.06$, if they include two Hickson compact groups with lower mass ($10^{11}~M_{\odot}<M_{200}<10^{12}~M_{\odot}$).
Our value for the sample including lower mass systems,  
$\alpha=0.92\pm0.05$, is larger than these values.
However, it is noted that the upper virial mass limit for the samples in \citet{rom17b} and \citet{man18}, $M_{200}=5\times 10^{14}~M_{\odot}$, is much smaller than the value in this study ($M_{200}=3\times 10^{15}~M_{\odot}$).

\subsection{The Mixed Formation Scenarios of UDGs}

There are three main scenarios proposed to explain the formation of UDGs in the literature: 1) the failed galaxies scenario, 2) the extended dwarf galaxies scenario, and 3) the interaction scenario \citep{amo16, van16, dic17, ben18}.

The first scenario is applicable to UDG progenitors undergoing early accretion to galaxy clusters.
\citet{yoz15} used $N$-body simulation to show how star formation in UDG progenitors could be quenched by ram-pressure stripping in cluster environments.
There are two variations of this scenario depending on the range of halo masses: a ``failed $L^{*}$ galaxies'' scenario and a ``failed dwarf galaxies'' scenario.
UDGs from ``failed $L^{*}$ galaxies'' have comparable halo masses to that of the MW ($M_{200}>10^{11}$ $M_{\odot}$), whereas ``failed dwarf galaxies'' have halo masses less than those of the Large Magellanic Cloud ($M_{200} < 10^{11}$ $M_{\odot}$).
Observationally, some Coma UDGs like DF44 or DFX1, which have red colors and massive dark matter (DM) halos ($M_{200}>10^{11}$ $M_{\odot}$), are consistent with the ``failed $L^{*}$ galaxies'' scenario  \citep{van16, van17}.
On the other hand, less massive UDGs with red colors such as DF17 or VCC 1287  ($M_{200}<10^{11}$ $M_{\odot}$) can be explained by the ``failed dwarf galaxies'' scenario \citep{bea16a, bea16b, pen16}.

In the ``extended dwarf galaxies'' scenario, UDG progenitors could be extended by internal processes. 
This scenario does not require high density environments or massive DM halos to produce UDGs.
\citet{amo16} and \citet{ron17} suggested that the high spin parameters of DM halos of UDGs could produce extended dwarf galaxies, and concluded that UDGs are the high-spin tails of normal dwarf galaxies.
In contrast, \citet{dic17} and \citet{cha18} suggested that strong gas outflows from stellar feedback could affect the gravitational potential of the DM halo and eventually extend the stellar content of UDG progenitors.
Blue isolated UDGs such as UGC 2162 \citep{tru17}, LSBG-750 \citep{gre18b}, and the UDGs in HCGs \citep{rom17b, spek18} have lower virial masses of $M_{\rm vir}\lesssim10^{11}$ $M_{\odot}$, which can be explained by this scenario.
Moreover, \citet{lei17} used HI observations and showed that UDGs in low-density environments tend to have relatively higher spin parameters than normal gas-rich galaxies.

In the third scenario, UDGs are created by tidal interactions with neighboring massive galaxies.
This scenario is supported by the presence of UDGs close to massive galaxies.
Several studies of UDG morphology provided observational evidence  supporting this scenario:
CenA-MM-Dw3 \citep{crn16}, DF4 near NGC 5485 \citep{mer16}, Scl-MM-Dw2 near NGC 253 \citep{tol16}, VLSB-A and VLSB-D in the Virgo cluster \citep{mih15, tol18}, NGC 2708-Dw1, and NGC 5631-Dw1 \citep{ben18}.
\citet{bau18} suggested that UDGs could be also generated via head-on collisions between two gas-rich galaxies which remove a large amount of gas in UDG progenitors.

Recent studies tend to suggest mixed formation scenarios to explain the diversity in the UDG populations in various environments by combining the individual scenarios \citep{Lee17, pap17, ala18, FeM18, Lim18, pan18}.
These studies revealed the presence of diverse UDGs through various methods: 1) the total number of GCs in UDGs, 2) optical spectroscopy, 3) SED fitting, and 4) HI observation.

First, the total number of GCs in UDGs has been used as a proxy to estimate the UDG halo mass.  
\citet{van17} presented that UDGs have, on average, a significantly larger number of GCs (by a factor of $\sim7$) than other galaxies with similar luminosity and stellar mass, using HST/ACS data of 16 Coma UDGs.
In their work, DF44 and DFX1 in Coma are good examples of UDGs with rich GC populations.
This implies that those UDGs are hosted by massive halos like the MW, but they failed to form as many stars as bright galaxies with similar dynamical masses.
From this, \citet{van17} concluded that their results support the failed $L^{*}$ galaxy scenario. 
However, \citet{amo18b} presented the opposite conclusion, using GCs in 54 Coma LSBs including 18 UDGs ($R_{\rm eff}>1.5$ \kpc).
They found that most of Coma UDGs in their sample have dwarf-like halos with $M_{\rm vir}<10^{11}$ $M_{\odot}$ with the 90\% confidence level, and only three UDGs are hosted by massive halos ($M_{\rm vir}>10^{11}$ $M_{\odot}$). 
Later, \citet{Lim18} studied the GC populations of 48 Coma UDGs.
They divided their UDG samples into two populations according to their GC specific frequency ($S_{N}$): high-$S_{N}$ UDGs and low-$S_{N}$ UDGs.
They revealed that high-$S_{N}$ UDGs (27 in total) have $M_{\rm halo}\sim10^{11}$ $M_{\odot}$ on average, whereas the remaining low-$S_{N}$ UDGs have significantly lower halo mass ($<10^{11}$ $M_{\odot}$).
This result is intermediate between \citet{van17} and \citet{amo18b} and implies that UDGs have two populations generated by multiple formation routes.
\citet{Lim18} concluded that UDGs have two populations formed by different routes. 
Likewise, \citet{tol18} applied a similar method to three Virgo UDGs.
They suggested that there are two types of UDGs: one is a smooth and DM-dominated system with a massive halo ($M_{200}\sim10^{12}$ $M_{\odot}$), and the other is a tidally perturbed system with significant rotation and a less massive halo ($M_{200}<10^{11}$ $M_{\odot}$). 

Second, optical spectroscopy has been used to study kinematics and stellar populations of UDGs.
\citet{ala18} obtained spectra of Coma UDGs using Keck/DEIMOS and derived the clustercentric radial velocities of the UDGs.
They suggested that Coma UDGs are divided into two different types in their velocity phase-space diagram (Fig 9. in their study).
One is `recent infall' UDGs ($\sim2$ $\rm Gyr$ ago) with high relative line-of-sight velocities and relatively bluer colors, and the other is early accreted `primordial' UDGs ($\sim8$ $\rm Gyr$ ago) with low relative line-of-sight velocities and redder colors.
\citet{FeM18} presented the results of stellar population analysis based on the same spectra as used in \citet{ala18}.
They derived the stellar parameters such as age, metallicity, and SFH, and showed that most UDGs have similar SFH, mass-age, and mass-metallicity relations to those of dwarf galaxies rather than to those of bright galaxies.
This supports that most UDGs have a dwarf-like nature.
However, a few UDGs like DF07 and DF44 in Coma show a `primordial' nature with low metallicities, old ages, and early quenching SFHs \citep{kad17, Gu18}.
\citet{chi19} derived kinematic and stellar parameters of 9 Coma UDGs/LSB dwarfs, using the MMT/Binospec spectrograph.
They suggested that UDGs and LSB dwarfs are of the same population and they have a wide range of age and metallicity due to a diversity of formation scenarios.
All these spectroscopic results imply that UDGs have multiple populations originated from different formation processes.

Third, SED fitting is also useful to reveal the multiple nature of UDGs.
Applying \texttt{prospector} (a fully Bayesian SED fitting package) to optical-NIR SEDs, \citet{pan18} studied two UDGs residing in different environments: VCC 1287 in the Virgo cluster, and DGSAT-I in the field environment.
They found that VCC 1287 shows a redder color ($g-i\sim0.7$), an older age ($\sim8$ $\rm Gyr$), a lower metallicity ($[Z/Z_{\odot}]\sim-1.0$), and a less extended SFH than DGSAT-I.
In contrast, DGSAT-I has a bluer color ($V-I\sim0.3$), a younger age ($\sim3$ $\rm Gyr$), a relatively metal-rich SED ($[Z/Z_{\odot}]\sim-0.6$), and an extended SFH. 
Although there have been not many studies of SEDs of UDGs, this implies that UDGs have a multiple formation scenario.

Fourth, recent HI observations of UDGs provided evidence of multiple populations of UDGs. 
\citet{pap17} presented the HI properties of four isolated UDGs.
Three UDGs (DGSAT I, R-127-1, and M161-1) are gas-deficient ($M_{\rm HI}/M_{*}\lesssim0.6$) and quiescent galaxies, while one UDG (SdI-2) is a gas-rich ($M_{\rm HI}/M_{*}>20$) dwarf galaxy. 
Considering that isolated UDGs such as UGC 2162 \citep{tru17} and SdI-1 \citep{bel17} have high gas fractions ($M_{\rm HI}/M_{*}>10$), UDGs in the low density environment can be distinguished as two types according to their gas fractions. 

In a similar context, our study of UDGs in massive clusters shows diverse types of UDGs.
A majority of UDGs are dwarf galaxies in terms of thier RDPs and dynamical masses.
Considering most UDGs have red colors and no star-forming features, they can be failed dwarf galaxies.
However, UDG-C22 in Abell 370 is so large ($R_{\rm eff,c}=6.16$ $\rm kpc$) and massive ($M_{200}>10^{11}$ $M_{\odot}$) that it can be considered as an example of a primodal failed $L^{*}$ galaxy.
Interestingly, UDG-C02 in Abell 370 can be an extended dwarf galaxy due to its remarkably blue color.
There are also a few tidally interacting UDGs.
This diversity of UDGs is consistently shown in the UDGs in other HFF clusters.
Thus, we conclude that UDGs in massive HFF clusters can be explained by multiple formation scenarios.

\section{Conclusion and Summary}
We used the HST archival images of Abell 370, a massive galaxy cluster in the HFF, to find and study UDGs and LSB dwarfs.
We investigated the properties of the UDGs and LSB dwarfs in Abell 370 in comparison with those in two other massive HFF clusters, Abell 2744 and Abell S1063.
The main results are summarized as follows.

\begin{enumerate}

	\item In the central and parallel HST fields of Abell 370, we found a total of 46 UDGs and 112 LSB dwarfs. There are 34 UDGs and 80 LSB dwarfs in the central field, and 12 UDGs and 32 LSB dwarfs in the parallel field.
	
	\item The CMDs of Abell 370 show that most UDGs are located in the red sequence of the cluster. This means that most UDGs are quiescent galaxies with no star formation. However, one UDG, UDG-C02, shows a much bluer color ($F814W-F105W=-0.07$), implying that this galaxy hosts a very young stellar population.
	
	\item Abell 370 UDGs mostly have exponential light profiles and round shapes. 
	Their structural parameters (S\'ersic indices ($n$) and axis ratios ($b/a$))  are, on average, similar to  those of UDGs in the Coma cluster and other HFF clusters.

	\item The RDPs of the galaxies in Abell 370 show a similar feature to those of Abell S1063 and Abell 2744. 
	The mean RDPs of UDGs and bright galaxies in the combined sample of the three HFF clusters show a significant discrepancy in the central region of the clusters.
    The profiles of UDGs  and LSB dwarfs show a flattening as the clustercentric distance decreases, while that of bright galaxies shows a continuous increase.
    This implies that UDGs and LSB dwarfs in the central regions of the clusters might have been tidally disrupted.
	
	\item We estimate the abundance of UDGs in Abell 370 from the RDP, obtaining $N(\rm UDG)=644\pm104$.
	This value is similar to those of Abell S1063 and Abell 2744.
	Combining our results on UDGs with those in the literature, we investigated the relation between the number of UDGs and the virial mass of their host systems. 
	This relation for the host mass range of $M_{200}>10^{13}~M_{\odot}$ is fitted very well by a power law with an index value close to one: $N({\rm UDG})\propto M_{200}^{0.99\pm0.05}$. 
	This  indicates that UDGs are formed with similar efficiency regardless of the virial mass of their host systems.

	\item Adding updated data of 13 UDG/LSB dwarfs, we derived a relation in the fundamental manifold method, ${\rm log}~V_{\rm est}=(0.90\pm0.03)\times{\rm log}~V_{\rm obs}+(0.04\pm0.03)$ with an RMS value of 0.16 dex, which is similar to \citet{zar17}'s.
	This shows that the fundamental manifold method can be applied to the UDG/LSB dwarf regime.
	
	\item We estimate the virial masses ($M_{200}$) of galaxies in Abell 370 with the fundamental manifold method, assuming that these galaxies are pressure-supported systems. Most UDGs have masses of $M_{200}=10^{10}-10^{11}$ $M_{\odot}$. However, a few UDGs such as UDG-C22 are more massive than $M_{200}>10^{11}$ $M_{\odot}$. This implies that most UDGs are hosted in dwarf halos, but a few of them are hosted in MW-like halos.
	
	\item UDGs and LSB dwarf galaxies do not show any significant distinction in their properties (size, luminosity, color, age, RDPs, and dynamical mass). This implies that UDGs are just a larger and more massive version of LSB dwarfs.
	
	\item In conclusion, our results support multiple formation scenarios of UDGs as suggested in the previous studies: Most UDGs have dwarf-like origins, while a few UDGs can be failed $L^{*}$ galaxies.
	
\end{enumerate}

Based on observations obtained with the NASA/ESA Hubble Space Telescope, retrieved from the Mikulski Archive for Space Telescopes (MAST) at the Space Telescope Science Institute (STScI).
STScI is operated by the Association of Universities for Research in Astronomy, Inc. under NASA contract NAS 5-26555.
This study was supported by the National Research Foundation of Korea (NRF) grant funded by the Korean Government (NRF-2019R1A2C2084019).
J.K. was supported by the Global Ph.D. Fellowship Program (NRF-2016H1A2A1907015).
We thank Brian S. Cho for helping to improve the English in the draft.
We would also like to thank the anonymous referee for giving very helpful comments and suggestions.

\clearpage

\begin{deluxetable}{ccc}
	\tabletypesize{\footnotesize}
	\setlength{\tabcolsep}{0.25in}
	\tablecaption{Physical Parameters of Abell 370 cluster\tablenotemark{\rm \footnotesize a}}
	\tablewidth{1000pt}
	\tablehead{ \colhead{Parameter} & \colhead{Value} & \colhead{References}}
	\startdata
	Redshift & $z=0.375$ & 1\\
	Distance Modulus & $(m-M)_{0}=41.44$ & 1\\
	Luminosity Distance & $1942$ $\rm{Mpc}$ & 1\\
	Angular Distance & $1028$ $\rm{Mpc}$ & 1\\
	Scale & $4.984$ $\rm{kpc}$ $\rm{arcsec^{-1}}$ & 1\\
	Age at Redshift & $9.366$ $\rm{Gyr}$ & 1\\
	Virial Radius & $r_{200}=8\farcm52\pm0\farcm36=2.55\pm0.11$ $\rm{Mpc}$ & 2 (weak+strong lensing analysis)\\
	Virial Mass & $M_{200}=(3.03\pm0.37)\times10^{15}$ $M_{\odot}$ & 2,3 (weak+strong lensing analysis)\\
	Foreground Reddening & $E(B-V)=0.028$ $\rm{mag}$ & 4\\
	Surface Brightness Dimming & $10~{\rm log}(1+z)=1.38$ \SBunit & 1\\
	\enddata
	\tablenotetext{}{\textbf{Notes.} References: (1) NASA/IPAC Extragalactic Database; (2) \citet{Umet11}; (3) \citet{broad08}; (4)  \citet{sch11}.}
	\tablenotetext{$\rm a$}{Based on cosmological parameters: $H_{0}=73$ $\rm{km}$ $\rm{s^{-1}}$ $\rm{Mpc^{-1}}$, $\Omega_{M}=0.27$, and  $\Omega_{\Lambda}=0.73$.}
	\label{tab_A370}
\end{deluxetable}

\begin{deluxetable}{cc}
	\tabletypesize{\footnotesize}
	\setlength{\tabcolsep}{0.50in}
	\tablecaption{Source Extractor Input Parameters Used in \textbf{Section 2.2}}
	\tablewidth{1000pt}
	\tablehead{ \colhead{Parameter name} & \colhead{Input configuration}}
	\startdata
	\texttt{DETECT\_MINAREA} & 20 pixels \\
	\texttt{DETECT\_THRESH} & 0.7 \\
	\texttt{ANALYSIS\_THRESH} & 0.7 \\
	\texttt{FILTER\_NAME} & tophat\_3.0\_3x3.conv \\
	\texttt{DEBLEND\_NTHRESH} & 32 \\
	\texttt{DEBLEND\_MINCONT} & 0.005 \\
	\texttt{PHOT\_AUTOPARAMS\tablenotemark{\rm \footnotesize a}} & 2.5, 3.5 \\
	\texttt{PHOT\_AUTOPARAMS\tablenotemark{\rm \footnotesize b}} & 1.25, 1.75 \\
	\texttt{BACK\_SIZE} & 32 pixels \\
	\texttt{BACKPHOTO\_TYPE} & LOCAL \\
	\enddata
	
	\tablenotetext{}{\textbf{Notes.}}
	\tablenotetext{$\rm a$}{These values are set to be default in SExtractor. We used these values to measure \texttt{MAG\_AUTO} for aperture magnitudes.}
	\tablenotetext{$\rm b$}{We set different values of \texttt{PHOT\_AUTOPARAMS} when measuring colors of sources, because smaller aperture radii can lead to higher signal-to-noise ratio.}
	
	\label{tab_SE}
\end{deluxetable}

\begin{deluxetable*}{cccc}
	\tabletypesize{\footnotesize}
	\setlength{\tabcolsep}{0.10in}
	\tablecaption{Numbers of Selected Sources in Each Step}
	\tablewidth{1000pt}
	\tablehead{\colhead{Sources} & \colhead{The Central Field} & \colhead{The Parallel Field} & \colhead{The XDF (HUDF09, HUDF12)}}
	\startdata
	\textbf{Step 1. SExtractor photometry} & & & \\
	Total detected sources & 18,315 & 15,998 & 12,014\\
	Galaxy candidates & 3,748 & 2,789 & 2,252\\
	Initial bright galaxies & 334 & 115 & 84\\ 
	Initial LSB galaxies (UDGs + LSB dwarfs) & 714 & 342 & 339\\ \hline
	\textbf{Step 2. GALFIT \& visual inspection} & & & \\
	Bright galaxy candidates & 315 & 106 & 80\\ 
	UDG candidates & 39 & 16 & 1\\
	LSB dwarf candidates & 87 & 35 & 18\\ \hline
	\textbf{Step 3. Color-magnitude relations} & & & \\
	Final bright galaxies & 298 & 93 & 56\\
	Final UDGs & 34 & 12 & 1\\
	Final LSB dwarfs & 80 & 32 & 10\\
	\enddata
	\label{tab_N}
\end{deluxetable*}

\begin{deluxetable*}{ccccccccc}
	\tabletypesize{\footnotesize}
	\setlength{\tabcolsep}{0.07in}
	\tablecaption{A Catalog of UDGs in Abell 370}
	\tablewidth{1000pt}
	\tablehead{\colhead{Name} & \colhead{R.A.} & \colhead{Decl.} &
		\colhead{$F814W$} & \colhead{$F814W-F105W$} & \colhead{\Reffc\tablenotemark{\rm \footnotesize a}} & \colhead{\absmue\tablenotemark{\rm \footnotesize b}} & 
		\colhead{$n$} & \colhead{$b/a$} \\
		\colhead{} & \colhead{(J2000)} & \colhead{(J2000)} &
		\colhead{} & \colhead{} & \colhead{[kpc]} & \colhead{[\SBunit]} & 
		\colhead{} & \colhead{}}	
	\startdata
    UDG-C01 & $39.94771$ & $-1.58373$ & $25.18\pm0.01$ & $0.22\pm0.01$ & $1.78\pm0.11$ & $23.96\pm0.08$ &$1.25\pm0.12$ & $0.81\pm0.03$ \\
    UDG-C02 & $39.94916$ & $-1.58248$ & $23.90\pm0.01$ & $-0.07\pm0.01$ & $2.82\pm0.02$ & $24.01\pm0.02$ &$0.53\pm0.02$ & $0.50\pm0.00$ \\
    UDG-C03 & $39.94958$ & $-1.58445$ & $25.17\pm0.01$ & $0.21\pm0.01$ & $1.64\pm0.08$ & $24.01\pm0.07$ &$0.80\pm0.08$ & $0.87\pm0.03$ \\
    UDG-C04 & $39.95016$ & $-1.57992$ & $26.23\pm0.03$ & $0.20\pm0.02$ & $1.94\pm0.19$ & $24.98\pm0.14$ &$1.20\pm0.20$ & $0.76\pm0.05$ \\
    UDG-C05 & $39.95378$ & $-1.57341$ & $24.81\pm0.01$ & $0.45\pm0.01$ & $1.90\pm0.06$ & $24.07\pm0.05$ &$0.62\pm0.04$ & $0.35\pm0.01$ \\	
	\enddata
	\tablenotetext{}{\textbf{Notes.}}
	\tablenotetext{$\rm a$}{\Reffc$~$$=R_{\rm eff}$$\sqrt{b/a}$ for the adopted distance scale of $4.984$ $\rm kpc$ $\rm arcsec^{-1}$.}
	\tablenotetext{$\rm b$}{\absmue = $\langle\mu\rangle_{\rm e} (r') - 10\times{\rm log} (1+z) - E(z) - K(z)$ at the redshift of $z=0.375$ \citep{Lee17}.
		\\(This table is available in its entirety in machine-readable form.)}
	\label{tab_UDG}
\end{deluxetable*}

\begin{deluxetable*}{ccccccccc}
	\tabletypesize{\footnotesize}
	\setlength{\tabcolsep}{0.07in}
	\tablecaption{A Catalog of LSB Dwarfs in Abell 370}
	\tablewidth{1000pt}
	\tablehead{\colhead{Name} & \colhead{R.A.} & \colhead{Decl.} &
		\colhead{$F814W$} & \colhead{$F814W-F105W$} & \colhead{\Reffc\tablenotemark{\rm \footnotesize a}} & \colhead{\absmue\tablenotemark{\rm \footnotesize b}} & 
		\colhead{$n$} & \colhead{$b/a$} \\
		\colhead{} & \colhead{(J2000)} & \colhead{(J2000)} &
		\colhead{} & \colhead{} & \colhead{[kpc]} & \colhead{[\SBunit]} & 
		\colhead{} & \colhead{}}	
	\startdata
    LDw-C01 & $39.94725$ & $-1.58327$ & $25.58\pm0.02$ & $0.06\pm0.01$ & $1.44\pm0.09$ & $24.02\pm0.08$ &$0.83\pm0.10$ & $0.63\pm0.03$ \\
    LDw-C02 & $39.95024$ & $-1.58280$ & $26.29\pm0.02$ & $-0.29\pm0.02$ & $1.09\pm0.17$ & $24.04\pm0.30$ &$2.45\pm0.52$ & $0.57\pm0.07$ \\
    LDw-C03 & $39.95126$ & $-1.57900$ & $25.72\pm0.02$ & $0.27\pm0.01$ & $1.39\pm0.10$ & $24.04\pm0.08$ &$1.00\pm0.13$ & $0.81\pm0.04$ \\
    LDw-C04 & $39.95405$ & $-1.57715$ & $25.87\pm0.02$ & $0.32\pm0.01$ & $1.22\pm0.05$ & $23.99\pm0.07$ &$0.59\pm0.04$ & $0.60\pm0.02$ \\
    LDw-C05 & $39.95444$ & $-1.56964$ & $26.53\pm0.03$ & $0.01\pm0.02$ & $1.28\pm0.14$ & $24.48\pm0.14$ &$1.39\pm0.23$ & $0.59\pm0.05$ \\
	\enddata
	\tablenotetext{}{\textbf{Notes.}}
	\tablenotetext{$\rm a$}{\Reffc$~$$=R_{\rm eff}$$\sqrt{b/a}$ for the adopted distance scale of $4.984$ $\rm kpc$ $\rm arcsec^{-1}$}
	\tablenotetext{$\rm b$}{\absmue = $\langle\mu\rangle_{\rm e} (r') - 10\times{\rm log} (1+z) - E(z) - K(z)$ for the redshift of $z=0.375$.
		\\(This table is available in its entirety in machine-readable form.)}
	\label{tab_LDw}
\end{deluxetable*}

\begin{deluxetable*}{ccccccccc}
	\tabletypesize{\footnotesize}
	\setlength{\tabcolsep}{0.20in}
	\tablecaption{Peak Values\tablenotemark{\rm \footnotesize a} of the De-reddened Colors of Each Galaxy Population}
	\tablewidth{2000pt}
	\tablehead{\colhead{Cluster} & \colhead{Galaxy population} & \colhead{$(F606W-F814W)_{0}$} & \colhead{$(F814W-F160W)_{0}$}}
	\startdata
	Abell 370 & Bright galaxies   & $0.88^{+0.01}_{-0.01}$ & $0.59^{+0.31}_{-0.01}$ \\
	& UDGs                        & $0.79^{+0.05}_{-0.06}$ & $0.39^{+0.02}_{-0.04}$ \\
	& LSB dwarfs                  & $0.65^{+0.08}_{-0.02}$ & $0.32^{+0.04}_{-0.03}$ \\ 
	\hline
	Abell S1063 & Bright galaxies & $0.86^{+0.01}_{-0.01}$ & $0.65^{+0.03}_{-0.03}$ \\
	& UDGs                        & $0.75^{+0.01}_{-0.11}$ & $0.39^{+0.03}_{-0.02}$ \\
	& LSB dwarfs                  & $0.72^{+0.01}_{-0.09}$ & $0.33^{+0.03}_{-0.05}$ \\
	\hline
	Abell 2744 & Bright galaxies  & $0.78^{+0.01}_{-0.01}$ & $0.56^{+0.09}_{-0.03}$ \\
	& UDGs                        & $0.74^{+0.01}_{-0.19}$ & $0.39^{+0.04}_{-0.03}$ \\
	& LSB dwarfs                  & $0.69^{+0.03}_{-0.10}$ & $0.21^{+0.14}_{-0.01}$ \\
	\enddata
	\label{tab_cc}
	\tablenotetext{}{\textbf{Notes.}}
	\tablenotetext{$\rm a$}{We obtained $1\sigma~(68.3\%)$ error values from bootstrap resampling.}
\end{deluxetable*}

\begin{deluxetable*}{cccc}
	\tabletypesize{\footnotesize}
	\setlength{\tabcolsep}{0.20in}
	\tablecaption{The Abundance of UDGs in Galaxy Groups and Clusters}
	\tablewidth{2000pt}
	\tablehead{\colhead{Host system} & \colhead{${\rm log}~M_{200}/M_{\odot}$ (host system)} & \colhead{$N({\rm UDG})^a$} & \colhead{References}}
	\startdata
	A370  & $15.48$ & $644\pm104$ & This study \\
	AS1063 & $15.43$ & $770\pm114$ & \citet{Lee17} \\
	A2744 & $15.34$ & $814\pm122$ & \citet{Lee17} \\
	Coma & $14.97$ & $193\pm14$ & \citet{yag16} \\
	A85 & $15.00$ & $189\pm21$ & \citet{vdB16} \\
	A119 & $14.88$ & $147\pm17$ & \citet{vdB16} \\
	A133 & $14.74$ & $110\pm17$ & \citet{vdB16} \\
	A780 & $14.79$ & $81\pm16$ & \citet{vdB16} \\
	A1781 & $13.90$ & $29\pm10$ & \citet{vdB16} \\
	A1795 & $14.72$ & $180\pm20$ & \citet{vdB16} \\
	A1991 & $14.28$ & $46\pm12$ & \citet{vdB16} \\
	MKW3S & $14.36$ & $51\pm11$ & \citet{vdB16} \\
	GAMA groups (bin 1) & $12.62$ & $0.3\pm0.6$ & \citet{vdB17} \\
	GAMA groups (bin 2) & $13.03$ & $1.4\pm0.5$ & \citet{vdB17} \\
	GAMA groups (bin 3) & $13.36$ & $3.3\pm0.8$ & \citet{vdB17} \\
	GAMA groups (bin 4) & $13.77$ & $7.5\pm2.4$ & \citet{vdB17} \\
	GAMA groups (bin 5) & $14.04$ & $45\pm13$ & \citet{vdB17} \\
	GAMA groups (bin 6) & $14.52$ & $64\pm17$ & \citet{vdB17} \\
	A168 & $14.23$ & $20\pm4.5$ & \citet{rom17a} \\
	UGC842 & $13.15$ & $6.0\pm2.5$ & \citet{rom17a} \\
	Fornax & $13.85$ & $33\pm9$ & \citet{ven17} \\
	HCG95 & $13.40$ & $7.0\pm4.7$ & \citet{shi17} \\
	HCG07 & $11.98$ & $3.0\pm1.7$ & \citet{rom17b} \\
	HCG25 & $11.83$ & $1.0\pm1.0$ & \citet{rom17b} \\
	HCG98 & $12.84$ & $2.0\pm1.4$ & \citet{rom17b} \\
	RXCJ1204.4+0154 & $13.46$ & $14\pm3.9$ & \citet{man18} \\
	A779 & $13.60$ & $20\pm4.6$ & \citet{man18} \\
	RXCJ1223.1+1037 & $13.30$ & $11\pm3.3$ & \citet{man18} \\
	MKW4S & $13.36$ & $5.0\pm2.2$ & \citet{man18} \\
	RXCJ1714.3+4341 & $12.78$ & $7.0\pm2.7$ & \citet{man18} \\
	A2634 & $14.42$ & $55\pm7.8$ & \citet{man18} \\
	A1177 & $13.58$ & $8.0\pm3.0$ & \citet{man18} \\
	A1314 & $13.88$ & $16\pm4.4$ & \citet{man18} \\
	\enddata
	\label{tab_ab}
	\tablenotetext{}{\textbf{Notes.}}
	\tablenotetext{$\rm a$}{UDGs were selected using the same selection criteria as in this study.}
\end{deluxetable*}
\clearpage


\begin{thebibliography}{}


    \bibitem[Abolfathi et al.(2018)]{abo18} Abolfathi, B., Aguado, D.~S., Aguilar, G., et al.\ 2018, \apjs, 235, 42


    \bibitem[Afanasiev et al.(2018)]{afa18} Afanasiev, A.~V., Chilingarian, I.~V., Mieske, S., et al.\ 2018, \mnras, 477, 4856


    \bibitem[Ahn et al.(2017)]{ahn17} Ahn, C.~P., Seth, A.~C., den Brok, M., et al.\ 2017, \apj, 839, 72


    \bibitem[Ahn et al.(2018)]{ahn18} Ahn, C.~P., Seth, A.~C., Cappellari, M., et al.\ 2018, \apj, 858, 102

	
	\bibitem[Alabi et al.(2018)]{ala18} Alabi, A., Ferr{\'e}-Mateu, A., Romanowsky, A.~J., et al.\ 2018, \mnras, 479, 3308  
	
	
	\bibitem[Amorisco \& Loeb(2016)]{amo16} Amorisco, N.~C., \& Loeb, A.\ 2016, \mnras, 459, L51 
	

	\bibitem[Amorisco et al.(2018)]{amo18b} Amorisco, N.~C., Monachesi, A., Agnello, A., \& White, S.~D.~M.\ 2018, \mnras, 475, 4235 
	
	
	\bibitem[Baushev(2018)]{bau18} Baushev, A.~N.\ 2018, \na, 60, 69 
	
	
	\bibitem[Beasley et al.(2016)]{bea16a} Beasley, M.~A., Romanowsky, A.~J., Pota, V., et al.\ 2016, \apjl, 819, L20 
	
	
	\bibitem[Beasley \& Trujillo(2016)]{bea16b} Beasley, M.~A., \& Trujillo, I.\ 2016, \apj, 830, 23 


    \bibitem[Bender et al.(1993)]{ben93} Bender, R., Burstein, D., \& Faber, S.~M.\ 1993, \apj, 411, 153	
    
	
	\bibitem[Bellazzini et al.(2017)]{bel17} Bellazzini, M., Belokurov, V., Magrini, L., et al.\ 2017, \mnras, 467, 3751 

	
	\bibitem[Bennet et al.(2018)]{ben18} Bennet, P., Sand, D.~J., Zaritsky, D., et al.\ 2018, \apj, 866, L11
 		
	
	\bibitem[Bertin \& Arnouts(1996)]{ber96} Bertin, E., \& Arnouts, S.\ 1996, \aaps, 117, 393 
	
	
	\bibitem[Broadhurst et al.(2008)]{broad08} Broadhurst, T., Umetsu, K., Medezinski, E., Oguri, M., \& Rephaeli, Y.\ 2008, \apjl, 685, L9 	
	
	
	\bibitem[Bruzual \& Charlot(2003)]{bc03} Bruzual, G., \& Charlot, S.\ 2003, \mnras, 344, 1000
	

	\bibitem[Caldwell \& Bothun(1987)]{cal87} Caldwell, N., \& Bothun, G.~D.\ 1987, \aj, 94, 1126
	

	\bibitem[Chabrier(2003)]{cha03} Chabrier, G.\ 2003, Publications of the Astronomical Society of the Pacific, 115, 763
	
	
	\bibitem[Chan et al.(2018)]{cha18} Chan, T.~K., Kere{\v s}, D., Wetzel, A., et al.\ 2018, \mnras, 478, 906 


    \bibitem[Chilingarian et al.(2011)]{chi11} Chilingarian, I.~V., Mieske, S., Hilker, M., et al.\ 2011, \mnras, 412, 1627	


    \bibitem[Chilingarian et al.(2019)]{chi19} Chilingarian, I.~V., Afanasiev, A.~V., Grishin, K.~A., et al.\ 2019, \apj, 884, 79


    \bibitem[Cohen et al.(2018)]{coh18} Cohen, Y., van Dokkum, P., Danieli, S., et al.\ 2018, \apj, 868, 96
	
	
	\bibitem[Collins et al.(2013)]{col13} Collins, M.~L.~M., Chapman, S.~C., Rich, R.~M., et al.\ 2013, \apj, 768, 172


    \bibitem[Collins et al.(2019)]{col19} Collins, M.~L.~M., Tollerud, E.~J., Rich, R.~M., et al.\ 2019, \mnras, 2833 	
 	
 	
 	\bibitem[Conselice et al.(2003)]{con03} Conselice, C.~J., Gallagher, J.~S., \& Wyse, R.~F.~G.\ 2003, \aj, 125, 66
 	
 	
 	\bibitem[Conselice(2018)]{con18} Conselice, C.~J.\ 2018, Research Notes of the American Astronomical Society, 2, 43
	
	
	\bibitem[Crnojevi{\'c} et al.(2016)]{crn16} Crnojevi{\'c}, D., Sand, D.~J., Spekkens, K., et al.\ 2016, \apj, 823, 19
	
	
	\bibitem[Di Cintio et al.(2017)]{dic17} Di Cintio, A., Brook, C.~B., Dutton, A.~A., et al.\ 2017, \mnras, 466, L1 
	
	
	\bibitem[Ferr{\'e}-Mateu et al.(2018)]{FeM18} Ferr{\'e}-Mateu, A., Alabi, A., Forbes, D.~A., et al.\ 2018, \mnras, 479, 4891 
	
	
	\bibitem[Geha et al.(2006)]{ge06} Geha, M., Blanton, M.~R., Masjedi, M., \& West, A.~A.\ 2006, \apj, 653, 240 
	

	\bibitem[Greco et al.(2018a)]{gre18a} Greco, J.~P., Greene, J.~E., Strauss, M.~A., et al.\ 2018a, \apj, 857, 104
	
	
	\bibitem[Greco et al.(2018b)]{gre18b} Greco, J.~P., Goulding, A.~D., Greene, J.~E., et al.\ 2018b, \apj, 866, 112

	\bibitem[Gu et al.(2018)]{Gu18} Gu, M., Conroy, C., Law, D., et al.\ 2018, \apj, 859, 37 


    \bibitem[Harris(2010)]{har10} Harris, W.~E.\ 2010, arXiv e-prints, arXiv:1012.3224


	\bibitem[Harris et al.(2017)]{har17} Harris, W.~E., Blakeslee, J.~P., \& Harris, G.~L.~H.\ 2017, \apj, 836, 67
	
	
	\bibitem[Illingworth et al.(2013)]{ill13} Illingworth, G.~D., Magee, D., Oesch, P.~A., et al.\ 2013, \apjs, 209, 6 
	

	\bibitem[Impey et al.(1988)]{imp88} Impey, C., Bothun, G., \& Malin, D.\ 1988, \apj, 330, 634

	
	\bibitem[Janssens et al.(2017)]{jan17} Janssens, S., Abraham, R., Brodie, J., et al.\ 2017, \apjl, 839, L17 
	

    \bibitem[Janssens et al.(2019)]{jan19} Janssens, S.~R., Abraham, R., Brodie, J., et al.\ 2019, \apj, 887, 92


	\bibitem[Kadowaki et al.(2017)]{kad17} Kadowaki, J., Zaritsky, D., \& Donnerstein, R.~L.\ 2017, \apjl, 838, L21 
	

	\bibitem[Koda et al.(2015)]{kod15} Koda, J., Yagi, M., Yamanoi, H., \& Komiyama, Y.\ 2015, \apjl, 807, L2 
	
	
	\bibitem[Kubo et al.(2007)]{kub07} Kubo, J.~M., Stebbins, A., Annis, J., et al.\ 2007, \apj, 671, 1466
	

	\bibitem[Lagattuta et al.(2017)]{Lag17} Lagattuta, D.~J., Richard, J., Cl{\'e}ment, B., et al.\ 2017, \mnras, 469, 3946 
	

	\bibitem[Lee et al.(2017)]{Lee17} Lee, M.~G., Kang, J., Lee, J.~H., \& Jang, I.~S.\ 2017, \apj, 844, 157 
	
	
	\bibitem[Leisman et al.(2017)]{lei17} Leisman, L., Haynes, M.~P., Janowiecki, S., et al.\ 2017, \apj, 842, 133 


	\bibitem[Lim et al.(2018)]{Lim18} Lim, S., Peng, E.~W., C{\^o}t{\'e}, P., et al.\ 2018, \apj, 862, 82  
	
	
	\bibitem[Lotz et al.(2017)]{Lot17} Lotz, J.~M., Koekemoer, A., Coe, D., et al.\ 2017, \apj, 837, 97 
	

	\bibitem[Mancera Pi{\~n}a et al.(2018)]{man18} Mancera Pi{\~n}a, P.~E., Peletier, R.~F., Aguerri, J.~A.~L., et al.\ 2018, \mnras, 481, 4381 


    \bibitem[Mancera Pi{\~n}a et al.(2019)]{man19} Mancera Pi{\~n}a, P.~E., Aguerri, J.~A.~L., Peletier, R.~F., et al.\ 2019, \mnras, 485, 1036

	
	\bibitem[Mart{\'{\i}}nez-Delgado et al.(2016)]{mar16} Mart{\'{\i}}nez-Delgado, D., L{\"a}sker, R., Sharina, M., et al.\ 2016, \aj, 151, 96 
	

    \bibitem[Mart{\'\i}n-Navarro et al.(2019)]{mar19} Mart{\'\i}n-Navarro, I., Romanowsky, A.~J., Brodie, J.~P., et al.\ 2019, \mnras, 484, 3425
	

    \bibitem[McConnachie(2012)]{Mc12} McConnachie, A.~W.\ 2012, \aj, 144, 4


    \bibitem[McLaughlin \& van der Marel(2005)]{Mc05} McLaughlin, D.~E., \& van der Marel, R.~P.\ 2005, \apjs, 161, 304


	\bibitem[Merritt et al.(2016)]{mer16} Merritt, A., van Dokkum, P., Danieli, S., et al.\ 2016, \apj, 833, 168 
	

    \bibitem[Mieske et al.(2008)]{mie08} Mieske, S., Hilker, M., Jord{\'a}n, A., et al.\ 2008, \aap, 487, 921

	
	\bibitem[Mihos et al.(2015)]{mih15} Mihos, J.~C., Durrell, P.~R., Ferrarese, L., et al.\ 2015, \apjl, 809, L21
	
	
	\bibitem[Mihos et al.(2017)]{mih17} Mihos, J.~C., Harding, P., Feldmeier, J.~J., et al.\ 2017, \apj, 834, 16
	

	\bibitem[M{\"u}ller et al.(2018)]{mul18} M{\"u}ller, O., Jerjen, H., \& Binggeli, B.\ 2018, \aap, 615, A105


	\bibitem[Mu{\~n}oz et al.(2015)]{mun15} Mu{\~n}oz, R.~P., Eigenthaler, P., Puzia, T.~H., et al.\ 2015, \apjl, 813, L15 
	
	
    \bibitem[Ordenes-Brice{\~n}o et al.(2018)]{ord18} Ordenes-Brice{\~n}o, Y., Eigenthaler, P., Taylor, M.~A., et al.\ 2018, \apj, 859, 52


    \bibitem[Pandya et al.(2016)]{pan16} Pandya, V., Mulchaey, J., \& Greene, J.~E.\ 2016, \apj, 819, 162
	
	
	\bibitem[Pandya et al.(2018)]{pan18} Pandya, V., Romanowsky, A.~J., Laine, S., et al.\ 2018, \apj, 858, 29 
	
	
	\bibitem[Papastergis et al.(2017)]{pap17} Papastergis, E., Adams, E.~A.~K., \& Romanowsky, A.~J.\ 2017, \aap, 601, L10 
	

    \bibitem[Peacock et al.(2010)]{pea10} Peacock, M.~B., Maccarone, T.~J., Knigge, C., et al.\ 2010, \mnras, 402, 803


	\bibitem[Peng et al.(2010)]{pen10} Peng, C.~Y., Ho, L.~C., Impey, C.~D., \& Rix, H.-W.\ 2010, \aj, 139, 2097 
	
	
	\bibitem[Peng \& Lim(2016)]{pen16} Peng, E.~W., \& Lim, S.\ 2016, \apjl, 822, L31 
	
	
	\bibitem[Pizagno et al.(2007)]{piz07} Pizagno, J., Prada, F., Weinberg, D.~H., et al.\ 2007, \aj, 134, 945 
	
	
	\bibitem[Rom{\'a}n \& Trujillo(2017a)]{rom17a} Rom{\'a}n, J., \& Trujillo, I.\ 2017a, \mnras, 468, 703 
	
	
	\bibitem[Rom{\'a}n \& Trujillo(2017b)]{rom17b} Rom{\'a}n, J., \& Trujillo, I.\ 2017b, \mnras, 468, 4039 


    \bibitem[Rom{\'a}n et al.(2019)]{rom19} Rom{\'a}n, J., Beasley, M.~A., Ruiz-Lara, T., et al.\ 2019, \mnras, 486, 823

	
	\bibitem[Rong et al.(2017)]{ron17} Rong, Y., Guo, Q., Gao, L., et al.\ 2017, \mnras, 470, 4231 
	

    \bibitem[Sandage \& Binggeli(1984)]{san84} Sandage, A., \& Binggeli, B.\ 1984, \aj, 89, 919
	

	\bibitem[Schlafly \& Finkbeiner(2011)]{sch11} Schlafly, E.~F., \& Finkbeiner, D.~P.\ 2011, \apj, 737, 103 
	

    \bibitem[Seth et al.(2014)]{set14} Seth, A.~C., van den Bosch, R., Mieske, S., et al.\ 2014, \nat, 513, 398

	
	\bibitem[Shi et al.(2017)]{shi17} Shi, D.~D., Zheng, X.~Z., Zhao, H.~B., et al.\ 2017, \apj, 846, 26 
	
	
	\bibitem[Shipley et al.(2018)]{shp18} Shipley, H.~V., Lange-Vagle, D., Marchesini, D., et al.\ 2018, \apjs, 235, 14 
	
	
	\bibitem[Sif{\'o}n et al.(2018)]{sif18} Sif{\'o}n, C., van der Burg, R.~F.~J., Hoekstra, H., Muzzin, A., \& Herbonnet, R.\ 2018, \mnras, 473, 3747 
	
	
	\bibitem[Simon \& Geha(2007)]{sige07} Simon, J.~D., \& Geha, M.\ 2007, \apj, 670, 313 
	
	
	\bibitem[Spekkens \& Karunakaran(2018)]{spek18} Spekkens, K., \& Karunakaran, A.\ 2018, \apj, 855, 28 
	
	
	\bibitem[Springob et al.(2007)]{spr07} Springob, C.~M., Masters, K.~L., Haynes, M.~P., Giovanelli, R., \& Marinoni, C.\ 2007, \apjs, 172, 599 


    \bibitem[Struble \& Rood(1999)]{str99} Struble, M.~F., \& Rood, H.~J.\ 1999, \apjs, 125, 35
	
	
	\bibitem[Toloba et al.(2016)]{tol16} Toloba, E., Sand, D.~J., Spekkens, K., et al.\ 2016, \apjl, 816, L5 
	
	
	\bibitem[Toloba et al.(2018)]{tol18} Toloba, E., Lim, S., Peng, E., et al.\ 2018, \apjl, 856, L31 
	

	\bibitem[Trujillo et al.(2017)]{tru17} Trujillo, I., Roman, J., Filho, M., \& S{\'a}nchez Almeida, J.\ 2017, \apj, 836, 191 
	

	\bibitem[Umetsu et al.(2011)]{Umet11} Umetsu, K., Broadhurst, T., Zitrin, A., Medezinski, E., \& Hsu, L.-Y.\ 2011, \apj, 729, 127 
	
	
	\bibitem[van der Burg et al.(2016)]{vdB16} van der Burg, R.~F.~J., Muzzin, A., \& Hoekstra, H.\ 2016, \aap, 590, A20 
	
	
	\bibitem[van der Burg et al.(2017)]{vdB17} van der Burg, R.~F.~J., Hoekstra, H., Muzzin, A., et al.\ 2017, \aap, 607, A79 
	
	
	\bibitem[van Dokkum et al.(2015)]{van15} van Dokkum, P.~G., Abraham, R., Merritt, A., et al.\ 2015, \apjl, 798, L45 
	

	\bibitem[van Dokkum et al.(2016)]{van16} van Dokkum, P., Abraham, R., Brodie, J., et al.\ 2016, \apjl, 828, L6 
	
	
	\bibitem[van Dokkum et al.(2017)]{van17} van Dokkum, P., Abraham, R., Romanowsky, A.~J., et al.\ 2017, \apjl, 844, L11 
	

    \bibitem[van Dokkum et al.(2019)]{van19} van Dokkum, P., Wasserman, A., Danieli, S., et al.\ 2019, \apj, 880, 91

	
	\bibitem[Venhola et al.(2017)]{ven17} Venhola, A., Peletier, R., Laurikainen, E., et al.\ 2017, \aap, 608, A142 
	

	\bibitem[Wittmann et al.(2017)]{wit17} Wittmann, C., Lisker, T., Ambachew Tilahun, L., et al.\ 2017, \mnras, 470, 1512 
	
	
	\bibitem[Yagi et al.(2016)]{yag16} Yagi, M., Koda, J., Komiyama, Y., \& Yamanoi, H.\ 2016, \apjs, 225, 11 
	
	
	\bibitem[Yozin \& Bekki(2015)]{yoz15} Yozin, C., \& Bekki, K.\ 2015, \mnras, 452, 937 
	
	
	\bibitem[Zaritsky et al.(2006a)]{zar06a} Zaritsky, D., Gonzalez, A.~H., \& Zabludoff, A.~I.\ 2006a, \apj, 638, 725 
	
	
	\bibitem[Zaritsky et al.(2006b)]{zar06b} Zaritsky, D., Gonzalez, A.~H., \& Zabludoff, A.~I.\ 2006b, \apjl, 642, L37 
	
	
	\bibitem[Zaritsky et al.(2008)]{zar08} Zaritsky, D., Zabludoff, A.~I., \& Gonzalez, A.~H.\ 2008, \apj, 682, 68 
	

    \bibitem[Zaritsky et al.(2012)]{zar12} Zaritsky, D., Colucci, J.~E., Pessev, P.~M., et al.\ 2012, \apj, 761, 93
    
    
    \bibitem[Zaritsky et al.(2013)]{zar13} Zaritsky, D., Colucci, J.~E., Pessev, P.~M., et al.\ 2013, \apj, 770, 121
    
    
    \bibitem[Zaritsky et al.(2014)]{zar14} Zaritsky, D., Colucci, J.~E., Pessev, P.~M., et al.\ 2014, \apj, 796, 71


	\bibitem[Zaritsky(2017)]{zar17} Zaritsky, D.\ 2017, \mnras, 464, L110 
	

\end{thebibliography}
\end{document}